\documentclass[manuscript,screen]{acmart}

\AtBeginDocument{%
  \providecommand\BibTeX{{%
    \normalfont B\kern-0.5em{\scshape i\kern-0.25em b}\kern-0.8em\TeX}}}

\setcopyright{acmcopyright}
\copyrightyear{2022}
\acmYear{2022}
\acmDOI{XXXXXXX.XXXXXXX}

\usepackage{amsmath,pstricks,color,dsfont}
\usepackage{graphicx}
\usepackage{psfrag,graphicx,epsfig}
\usepackage{multirow}
\usepackage{url}
\usepackage{textcomp}
\usepackage{algorithm,algorithmic,xspace}
\usepackage[algo2e,vlined,linesnumbered,boxruled]{algorithm2e}

\let\oldnl\nl
\newcommand{\nonl}{\renewcommand{\nl}{\let\nl\oldnl}}

\allowdisplaybreaks

\newcommand{\bremark}{\begin{remark} \begin{rm} }
\newcommand{\eremark}{ \end{rm} \rule{1mm}{2mm}
\end{remark} }
\newcommand{\btheorem}{\begin{theorem} \begin{rm} }
\newcommand{\etheorem}{ \end{rm} \rule{1mm}{2mm}
\end{theorem} }
\newcommand{\blemma}{\begin{lemma} \begin{rm} }
\newcommand{\elemma}{ \end{rm} \rule{1mm}{2mm}
\end{lemma} }
\newcommand{\bcorollary}{\begin{corollary} \begin{rm} }
\newcommand{\ecorollary}{ \end{rm} \rule{1mm}{2mm}
\end{corollary} }
\newcommand{\bdefinition}{\begin{definition}\begin{rm} }
\newcommand{\edefinition}{ \end{rm} \rule{1mm}{2mm}
\end{definition} }
\newcommand{\bproposition}{\begin{proposition} \begin{rm} }
\newcommand{\eproposition}{ \end{rm} \rule{1mm}{2mm}
\end{proposition} }
\newcommand{\bexample}{\begin{example} \begin{rm} }
\newcommand{\eexample}{ \end{rm} \rule{1mm}{2mm}
\end{example} }
\newcommand{\basm}{\begin{assumption} \begin{rm}}
\newcommand{\easm}{\end{rm}
\end{assumption}}

\newtheorem{theorem}{\bf Theorem}[section]
\newtheorem{lemma}{\bf Lemma}[section]
\newtheorem{definition}{\bf Definition}[section]
\newtheorem{remark}{\bf Remark}[section]
\newtheorem{corollary}{\bf Corollary}[section]
\newtheorem{proposition}{\bf Proposition}[section]
\newtheorem{example}{\bf Example}[section]
\newtheorem{assumption}{\bf Assumption}[section]

\newcommand\oprocendsymbol{\hbox{$\bullet$}}
\newcommand\oprocend{\relax\ifmmode\else\unskip\hfill\fi\oprocendsymbol}

\begin{document}







\title{Privacy-preserving Decentralized Federated Learning over Time-varying Communication Graph}

\author{Yang Lu}
\affiliation{%
  \institution{Lancaster University}
  \streetaddress{InfoLab21}
  \city{Lancaster}
  \country{UK}}
  \postcode{LA1 4WA}
\email{y.lu44@lancaster.ac.uk}

\author{Zhengxin Yu}
\affiliation{%
  \institution{Lancaster University}
  \streetaddress{InfoLab21}
  \city{Lancaster}
  \country{UK}}
  \postcode{LA1 4WA}
\email{z.yu8@lancaster.ac.uk}

\author{Neeraj Suri}
\affiliation{%
  \institution{Lancaster University}
  \streetaddress{InfoLab21}
  \city{Lancaster}
  \country{UK}}
  \postcode{LA1 4WA}
\email{neeraj.suri@lancaster.ac.uk}

\begin{abstract}
Establishing how a set of learners can provide privacy-preserving federated learning in a fully decentralized (peer-to-peer, no coordinator) manner is an open problem. We propose the first privacy-preserving consensus-based algorithm for the distributed learners to achieve decentralized global model aggregation in an environment of high mobility, where the communication graph between the learners may vary between successive rounds of model aggregation. In particular, in each round of global model aggregation, the Metropolis-Hastings method \cite{XIAO200733} is applied to update the weighted adjacency matrix based on the current communication topology. In addition, the Shamir's secret sharing scheme \cite{Shamir} is integrated to facilitate privacy in reaching consensus of the global model. The paper establishes the correctness and privacy properties of the proposed algorithm. The computational efficiency is evaluated by a simulation built on a federated learning framework with a real-word dataset.
\end{abstract}



\maketitle

\section{Introduction}

\subsection{Background and Motivation}

Federated learning is a collaborative machine learning technique providing privacy preservation of the individual learners' local training data \cite{pmlr-v54-mcmahan17a,DBLP:journals/corr/KonecnyMRR16}. Each learner downloads the current global model from a centralized server, updates it by incorporating its local training data, and then sends the updated model back to the server. The server then aggregates the local models of all the individual learners to update the global model. Thus, only local training models can be observed during the training process, while raw training data do not leave their owners' devices. Given this significant feature of privacy preservation, federated learning has been applied to a wide range of applications, including wireless communications \cite{SN-HSD-JHR:2019}, autonomous driving \cite{elbir2020federated}, multi-access edge computing \cite{9181482}, smart manufacturing \cite{su12020492}, and healthcare \cite{JX-BSG-CS-PW-JB-FW:2021}.

The traditional federated learning paradigm has two major issues. First, it requires a centralized server such that it is connected to all the local learners. In some scenarios, the learners are geographically dispersed over a large area and may lack such a connect-to-all server. In addition, the paradigm is not robust, since if the single centralized server fails, then the whole learning task cannot proceed. Second, the local training models are directly uploaded to the centralized server. As has been recently pointed out \cite{10.1145/2810103.2813677,DBLP:conf/sp/NasrSH19}, it is possible that private local training data can be reconstructed from local training models via model inference or inversion attacks. The above two issues necessitate new mechanisms that can achieve federated learning in a decentralized and privacy-preserving manner.

\subsection{Related Works}

Multiple recent works have addressed the issue with fixed centralized server. Based on the technologies in achieving model aggregation, these works can be mainly categorized into two classes. The first class of works dynamically selects a learner to take the role of the centralized server \cite{DBLP:journals/corr/abs-1905-06731,DBLP:journals/corr/abs-2108-00365,9359135}. Informally, for each round of model updates, a learner is first selected, either randomly or by following specific rules. All the other learners send their local models, possibly relayed via in-between learners, to the selected learner, who then performs model aggregation to update the global model. This approach requires coordination between the learners for aggregation learner selection for each round of model update. Another class of works adopts consensus-based algorithms, where the learners iteratively update their local models to reach consensus on the desired global model \cite{9086196,DBLP:journals/corr/abs-1901-11173,8950073}. At each iteration, the learners exchange their local models only with their one-hop neighbors. In contrast to the first approach, the consensus-based approach does not require coordination between the learners and hence is easier for practical implementation. However, all these works only consider a fixed communication topology and not applicable to an environment of high mobility where the communication topology may change between successive rounds of model aggregation. In addition, all the aforementioned works directly exchange local models between the learners and thus still suffer from model inference and inversion attacks.

In this work, we develop the first privacy-preserving consensus-based decentralized federated learning algorithm that considers mobility. This is closely related to the problem of privacy-preserving consensus, where the target is to protect the privacy of the participants' initial states in the process of reaching consensus.

Existing works on privacy-preserving consensus can be categorized into four classes. 

The first class of works uses perturbation-based approaches. An important branch of works in this class uses the technique of differential privacy \cite{ZH-SM:2012,8486684,9133180,NOZARI2017221}. Differentially private schemes add random perturbations into individuals' private
data such that the participation of an individual cannot be inferred via perturbed data by an adversary with access to arbitrary auxiliary information \cite{Dwork}. Due to the usage of persistent random noises, there is a fundamental trade-off between privacy and utility \cite{QG-PV:2014,YL-MZ:2019ARC}. The very recent work \cite{9484437} proposed a different perturbation-based approach, which used the number of iterations between two
learners being in the same group (called gap therein) to control communication patterns among the learners. This approach also has a fundamental trade-off between privacy and utility.

The second class of works obfuscates exchanged data by adding decaying or correlated noises, which can guarantee consensus accuracy \cite{7465717,6669251,8356738,8619133,8430960}. This approach ensures that the private data cannot be uniquely determined. However, it still causes privacy leakage in the sense of information entropy of the private data, and the level of privacy leakage is determined by the magnitude of the noises \cite{8356738}. 

The third class of works adopts the technique of homomorphic encryption \cite{6375935,KK-TF:2015,YS-KG-AA-GJP-SAS-MS-PT:2016,YL-MZ:2018Automatica}. Informally speaking, homomorphic encryption allows certain algebraic operations to be carried out on ciphertexts, thus generating an encrypted result which, when decrypted, matches the result of operations
performed on plaintexts \cite{XY-RP-EB:2014}. Existing homomorphic encryption-based works require the existence of a centralized third party to carry out aggregation over ciphertexts. Hence, they are not applicable to the decentralized setting.

The fourth class of works leverages state decomposition to achieve privacy-preserving consensus in a decentralized setting \cite{8601358,8657789}. In this approach, a scalar step size shared between two neighboring agents is constructed as a product of two scalar numbers, each randomly generated by one of the two agents and kept unknown to the other one. During the consensus algorithm, the agents exchange the product of their states and the randomly generated step size splits. Without knowing their step size splits, one agent cannot determine the values of the states of its neighbors. However, to guarantee convergence of the underlying consensus algorithm, the step size splits have to be restricted in a small interval. This will cause privacy degradation, as one can have a good estimate of the value of an agent's state by knowing the admissible interval and observing the product of the state and the step size split.

\subsection*{Positioning our Research}

To overcome the above limitations of existing works, we propose a new algorithm which integrates Shamir's secret sharing (SSS) to achieve privacy-preserving consensus-based decentralized federated learning. Informally speaking, SSS distributes a secret among a group of participants, each of whom is allocated a share of the secret. As established by Shamir \cite{Shamir}, the secret can be reconstructed only when a sufficient number of shares are combined together, while a smaller number of shares contain no information of the secret. This technique has been widely applied to secure multiparty computation (SMC) on complete graphs \cite{RC-ID-JBN:2015}, where each participant can communicate with each other participant. Roughly speaking, each participant sends one share of its secret to each other participant. Each participant then computes an aggregation of the shares it receives from all the other participants. When a sufficient number of aggregated results are combined, the desired aggregation of the secrets of all the participants can be reconstructed. While these approaches work well in fully connected graphs, most real-world applications entail sparse graphs, e.g., optimal resource allocation in power systems \cite{7035105}, multi-robot formation control \cite{7487747}, and distributed environmental monitoring \cite{6630952}. In addition, in an environment of high mobility, the communication topology may change over time. For SMC over time-varying sparse graphs, where, at each round of computation, each participant can only communicate with its current neighbors, the above mechanisms cannot be applied. Few research has been conducted to SMC over sparse graphs. An exception is the recent work \cite{8903166}, which applied SSS to achieve privacy-preserving average consensus over sparse graphs. The work \cite{8903166} has three major limitations. First, the approach of \cite{8903166} needs to randomly activate one learner at each iteration, which requires coordination between the learners. Second, the approach of \cite{8903166} can only deal with the case where each learner has at least two neighbors and the three learners form a fully connected graph. Third, rigorous correctness and privacy analysis are absent in \cite{8903166}. The paucity of SMC research on time-varying sparse graphs motivates our work to establish fundamental results therein.

\subsection{Overview of Approach and Contributions}

This paper considers the problem of privacy-preserving decentralized federated learning over a time-varying communication graph. Specifically, we consider the case where the global training model is updated as a weighted average of the learners' local training models, and an average consensus algorithm is adopted to achieve decentralized aggregation. In each round of model aggregation, the Metropolis-Hastings method \cite{XIAO200733} is applied to update the weighted adjacency matrix based on the current communication topology to ensure convergence of average consensus. To protect the privacy of local training models against semi-honest learners, the learners use the Shamir's secret sharing scheme \cite{Shamir} to distribute their local models to their one-hop neighbors. Upon receiving the shares from its neighbors, each learner updates its model by inputting the sum of the shares it holds to the consensus algorithm. The contributions of our work are fourfold.
\begin{itemize}
    \item First, the proposed algorithm is the first that can achieve federated learning over a time-varying communication graph in a fully decentralized (without any coordination between the learners) and provably privacy-preserving manner.
    \item Second, in terms of privacy-preserving consensus, the proposed algorithm, for the first time, simultaneously achieves the following properties: (i) applicable to an arbitrary undirected connected communication graph without the need of a third party; (ii) perfect consensus (model aggregation) accuracy; (iii) no additional privacy leakage beyond the learners' own inputs (the local training models) and outputs (the updated global models); (iv) no privacy-convergence tradeoff.
    \item Third, the correctness and privacy properties of the proposed algorithm are rigorously analyzed. In particular, the correctness analysis addresses new challenges brought by signed real-valued models and termination of consensus iteration, and the privacy analysis addresses new challenges in potential additional privacy leakage caused by consensus process and time-varying communication topology. Please refer to Section \ref{subsec:new challenges} for detailed discussions.
    \item Fourth, the correctness and computational efficiency of the proposed algorithm are demonstrated by a simulation on a federated learning framework using a real-world dataset.
\end{itemize}

\subsection{Organization}

The rest of this paper is organized as follows. Section \ref{sec:problem statement} introduces the problem statement. Section \ref{sec:technical preliminaries} provides some necessary technical preliminaries. New challenges in algorithm design and analysis are identified in Section \ref{sec:new challenges}. The proposed algorithm is detailed in Section \ref{sec:algorithm design}. Its correctness and privacy properties are analyzed in Section \ref{sec:analysis}. In Section \ref{sec:simulation}, case studies are presented to test the performance of the proposed algorithm. Conclusions and future works are found in Section \ref{sec:conclusion}.



\section{Problem Statement}\label{sec:problem statement}

In this section, we first review the framework of centralized federated learning. Next, we formulate the problem of decentralized federated learning over a time-varying communication graph and identify its privacy issue. Subsequently, we introduce the adopted attacker model and privacy definition. Finally, we clarify the objectives of the paper.



\subsection{Centralized Federated Learning}

Consider a set of $N$ learners $\mathcal{V}\triangleq\{1,\cdots,N\}$. Each learner $i$ holds a set of $m_i\in\mathbb{N}$ local data samples, denoted by $D_i$. The learners aim to collaboratively train a common global model $\theta\in\mathbb{R}^n$ over all the $D_i$'s. In federated learning, for the purpose of preserving privacy of individual $D_i$'s, in each round $t$ of model update, each learner $i$ first trains a local model $\theta_i^{(t)}\in\mathbb{R}^n$ over $D_i$. This can be expressed as
\begin{align}
\label{local model training}
    \theta_i^{(t)}=\mathcal{F}_i(\theta_i^{(t,0)},D_i),
\end{align}
where $\theta_i^{(t,0)}\in\mathbb{R}^n$ is the initial model for learner $i$'s local training in round $t$, and $\mathcal{F}_i$ is its local training algorithm, e.g., a stochastic gradient descent-based algorithm \cite{HBM-EM-DR-SH-BAYA:2017}.

The global model $\theta^{(t)}\in\mathbb{R}^n$ is derived by performing a weighted aggregation over all the $\theta_i^{(t)}$'s as
\begin{align}
\label{aggregation}
    \theta^{(t)}=\sum_{i\in\mathcal{V}}w_i\theta_i^{(t)},
\end{align}
where $w_i>0$ is the weight on $\theta_i^{(t)}$. A popular choice of $w_i$ is given by $w_i=\frac{m_i}{m}$ with $m=\sum_{i\in\mathcal{V}}m_i$, i.e., $w_i$ is the proportion of learner $i$'s training data in the overall training data. Notice that in an execution of Eq. \eqref{aggregation}, only the local training models $\theta_i^{(t)}$'s can be observed, while the raw training data never leave their owners' devices.

\begin{figure}[!ht]
\begin{center}
\includegraphics[width=1\linewidth]{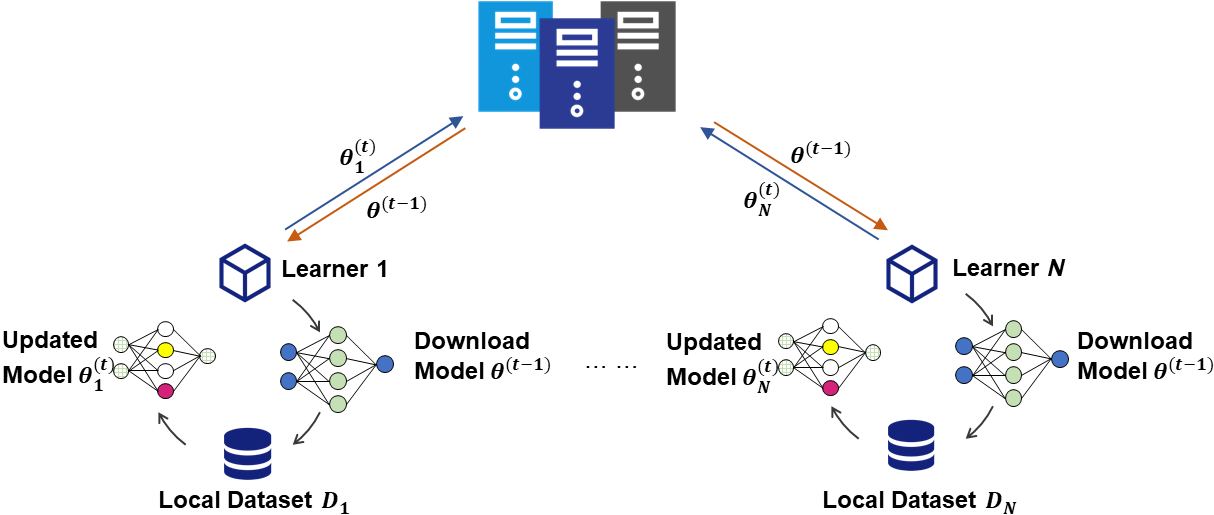}
\caption{Centralized federated learning.}
\label{centralized_FL}
\end{center}
\end{figure}

In the centralized setting, as shown by Fig. \ref{centralized_FL}, each learner $i$ uploads $\theta_i^{(t)}$ to a centralized server. Upon receiving the local models from all the learners, the centralized server updates the global model $\theta^{(t)}$ by Eq. \eqref{aggregation} and sends $\theta^{(t)}$ to all the learners. Each learner $i$ then sets $\theta_i^{(t+1,0)}=\theta^{(t)}$ and $t\leftarrow t+1$, and progresses to Eq. \eqref{local model training} for the next round of local training.

\subsection{Decentralized Federated Learning over A Time-Varying Communication Graph}

\begin{figure}[!ht]
\begin{center}
\includegraphics[width=1\linewidth]{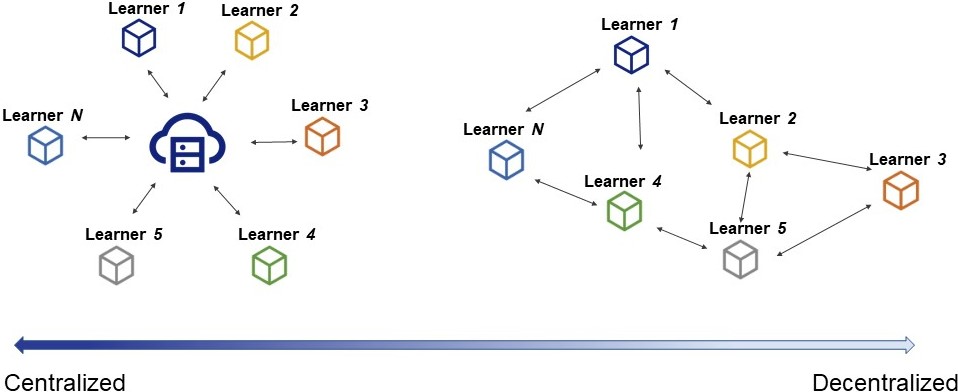}
\caption{Centralized aggregation vs decentralized aggregation.}
\label{decentralized_FL}
\end{center}
\end{figure}

\textbf{\emph{Decentralized model aggregation.}} As mentioned, in some scenarios, especially when the learners are geographically dispersed over a large area, there may not exist a centralized server that is connected to all the learners; please see Fig. \ref{decentralized_FL} as an illustration. In such cases, the learners need to carry out the model aggregation Eq. \eqref{aggregation} in a decentralized manner over the underlying communication graph between them.

\begin{figure}[!ht]
\begin{center}
\includegraphics[width=1\linewidth]{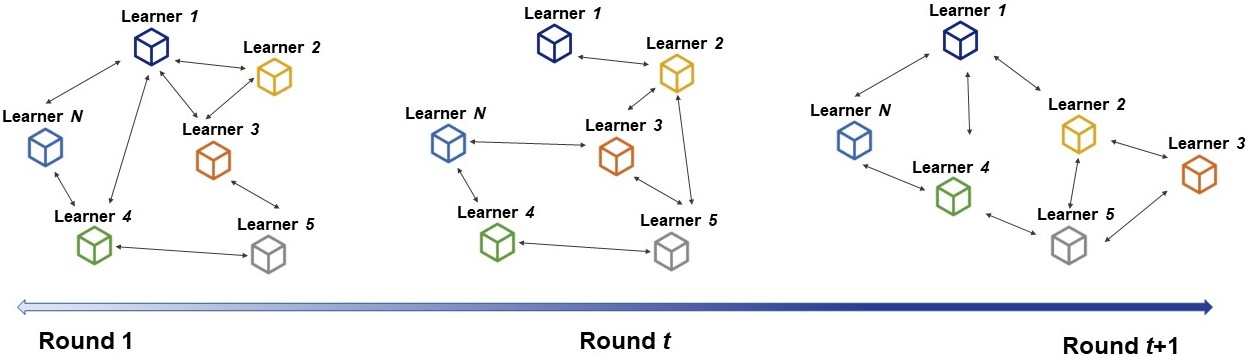}
\caption{Communication topology change between successive rounds of model aggregation.}
\label{topology_change}
\end{center}
\end{figure}

\noindent\textbf{\emph{Time-varying communication graph.}} In an environment of high mobility, the communication topology between the learners may vary between successive rounds of model aggregation as depicted in Fig. \ref{topology_change}.

Denote $\mathcal{G}^{(t)}=(\mathcal{V},\mathcal{E}^{(t)})$ as the communication graph between the learners during the $t$-th round of model aggregation, where $\mathcal{E}^{(t)}\subseteq\mathcal{V}\times\mathcal{V}$ is the set of communication links such that $(i,j)^{(t)}\in\mathcal{E}^{(t)}$ if and only if learner $i$ can receive messages from learner $j$ during the $t$-th round of model aggregation. Denote $\mathcal{N}_i^{(t)}\subseteq\mathcal{V}$ as the set of neighbors of learner $i$ in $\mathcal{G}^{(t)}$, i.e., $\mathcal{N}_i^{(t)}=\{j\in\mathcal{V}\setminus\{i\}:(i,j)^{(t)}\in\mathcal{E}^{(t)}\}$. Denote $\bar{\mathcal{N}}_i^{(t)}=\mathcal{N}_i^{(t)}\cup\{i\}$. Throughout the paper, we have the following mild assumption on $\mathcal{G}^{(t)}$.

\begin{assumption}
\label{asm:communication graph}
For any $t\in\mathbb{N}$, $\mathcal{G}^{(t)}$ is undirected, connected, and time invariant within the $t$-th round of model aggregation.
\end{assumption}


\textbf{\emph{Privacy issue.}} During each round $t$ of model aggregation, for each learner $i$, its local model $\theta_i^{(t)}$ must be kept private to itself, as breach of $\theta_i^{(t)}$ may enable an attacker to reconstruct learner $i$'s local training data by inference or inversion attacks.

\subsection{Attacker Model}

We consider the semi-honest attacker model, i.e., an adversarial learner correctly follows the designed algorithm but attempts to use its received data to infer others' private data (\cite{Hazay}, pp-20). The semi-honest attacker model has been widely used in various applications, e.g., privacy-preserving linear programming, dataset process and consensus \cite{JD-FK:2011,MJF-KN-BP,ZH-SM:2012}. We assume that the communication links between the learners are secure\footnote{Secure communication links can be enforced by cryptographic technologies such as encryption schemes.}.

\subsection{Privacy Definition}

As discussed above, our concerned problem is how all the learners can collaboratively compute the correct global model $\theta^{(t)}$ without disclosing their local models $\theta_i^{(t)}$'s to other learners. This is a secure multiparty computation (SMC) problem. Perfect secrecy, which will be adopted in this paper, is a standard privacy notion for SMC. Roughly speaking, an algorithm provides perfect secrecy if, after executing the algorithm, the adversarial entities only know their own inputs and outputs, but do not know anything beyond them, even if they have unlimited computing power \cite{Shannon}. It is worth noting that, unlike perturbation-related privacy notions, e.g., differential privacy, perfect secrecy does not induce the issue of utility-privacy tradeoff.

We next provide the formal definition of perfect secrecy in the general context of SMC, where, given a set of entities $\mathcal{V}$, each entity $i\in\mathcal{V}$ has a secret input $x_i$ and aims to compute the value of $f_i(\{x_j\}_{j\in\mathcal{V}})$. To do that, we need to introduce the notions of \emph{perfect indistinguishability} and \emph{view}. First, the following definition states that two distributions are perfectly indistinguishable if they follow the same distribution.

\begin{definition}[\cite{RC-ID-JBN:2015}]
\label{def: perfect indistinguishability}
Let $\mathcal{X}=\{\mathcal{X}(\kappa)\}_{\kappa\in\mathbb{N}}$ and $\mathcal{Y}=\{\mathcal{Y}(\kappa)\}_{\kappa\in\mathbb{N}}$ be two distribution ensembles, where, for each $\kappa\in\mathbb{N}$, $\mathcal{X}(\kappa)$ and $\mathcal{Y}(\kappa)$ are two random variables with the same probability space and the same range $R(\kappa)$. We say that $\mathcal{X}$ and $\mathcal{Y}$ are perfectly indistinguishable, denoted by $\mathcal{X}\mathop  \equiv \limits^p\mathcal{Y}$, if the following holds
\begin{align*}
\sum_{r\in R(\kappa)}|\Pr[\mathcal{X}(\kappa)=r]-\Pr[\mathcal{Y}(\kappa)=r]|=0,\;\;\forall \kappa\in\mathbb{N}.
\end{align*}
\end{definition}

Next, we introduce the notion of \emph{view}. Informally, the view of an entity is the set of all the messages the entity can see after the execution of the algorithm.

\begin{definition}[\cite{RC-ID-JBN:2015,OG:2004}]
\label{def: view}
Let $\Pi$ be an algorithm for computing $f=\{f_i\}_{i\in\mathcal{V}}$. For an execution of $\Pi$ on a joint input $x=\{x_i\}_{i\in\mathcal{V}}$, the view of entity $i$, denoted by ${\rm VIEW}_i^\Pi(x)$, is ${\rm VIEW}_i^\Pi(x)\triangleq\{x_i,m_1^i,\cdots,m_{t_i}^i\}$, where
$t_i$ is the total number of messages received by entity $i$, and for each $\ell\in\{1,\cdots,t_i\}$, $m_\ell^i$ is the $\ell$-th message it receives.
\end{definition}

This provides the basis to define perfect secrecy.


\begin{definition}[\cite{RC-ID-JBN:2015}]
\label{def: perfect secrecy}
Let $\Pi$ be an algorithm for computing $f=\{f_i\}_{i\in\mathcal{V}}$. Given a joint input $x=\{x_i\}_{i\in\mathcal{V}}$, denote the joint view of the entities in a set $\mathcal{I}\subseteq\mathcal{V}$ by ${\rm VIEW}_{\mathcal{I}}^\Pi(x)$. Let $\mathcal{A}$ be the set of adversarial learners. We say that $\Pi$ provides perfect secrecy against $\mathcal{A}$ if there exists a probabilistic polynomial-time algorithm $S$, such that for any admissible $x$, it holds that
\begin{align}
\label{eq perfect secrecy}
S(\mathcal{A},\{x_i\}_{i\in \mathcal{A}},\{f_i\}_{i\in \mathcal{A}})\mathop  \equiv \limits^p{\rm VIEW}_{\mathcal{A}}^\Pi(x).
\end{align}
\end{definition}

The condition Eq. \eqref{eq perfect secrecy} implies that whatever can been seen by $\mathcal{A}$ after the execution of $\Pi$ can be simulated by an algorithm $S$ using only $\mathcal{A}$'s own inputs and outputs, and $\mathcal{A}$ cannot distinguish $S(\mathcal{A},\{x_i\}_{i\in \mathcal{A}},\{f_i\}_{i\in \mathcal{A}})$ and ${\rm VIEW}_{\mathcal{A}}^\Pi(x)$ even if it has unlimited computing power. In other words, the execution of $\Pi$ does not provide $\mathcal{A}$ any additional information beyond what it must know, i.e., $\mathcal{A}$'s own inputs and outputs.

\subsection{Design Objectives}\label{objectives}

In this paper, we aim to design a privacy-preserving decentralized algorithm for the model aggregation Eq. \eqref{aggregation} over a time-varying sparse communication graph satisfying Assumption \ref{asm:communication graph}, such that the following properties are simultaneously guaranteed:
\begin{itemize}
\item Correctness: For every round $t\in\mathbb{N}$, all the learners derive the correct global model $\theta^{(t)}$ given by Eq. \eqref{aggregation}.

\item Privacy: The proposed algorithm protects the privacy of benign learners' local models $\theta_i^{(t)}$'s against semi-honest learners in the sense of perfect secrecy.
\end{itemize}

\section{Technical Preliminaries}\label{sec:technical preliminaries}

In this paper, we achieve the objectives stated in Section \ref{objectives} by integrating average consensus and Shamir's secret sharing. This section provides necessary technical preliminaries of the two techniques.

\subsection{Consensus-based Decentralized Model Aggregation}

Average consensus is an effective method to achieve decentralized aggregation over sparse communication graphs. This subsection first provides preliminaries on average consensus-based decentralized model aggregation, then introduces the Metropolis-Hastings method to deal with time-varying communication graphs. More detailed discussions can be found in \cite{XIAO200733,1272421,6854643}.

\noindent\textbf{\emph{Average consensus.}} Roughly speaking, this method enables a set of entities over a sparse connected communication graph, each with an initial state, to iteratively interact with their neighbors and update their states, such that all the entities' states will asymptotically converge to the average of their initial states.

To apply the average consensus method, for each round $t$ of model aggregation, the communication graph $\mathcal{G}^{(t)}$ needs to be equipped with a weighted adjacency matrix $A^{(t)}=[a_{ij}^{(t)}]\in\mathbb{R}^{N\times N}$ such that $a_{ij}^{(t)}>0$ if $(i,j)^{(t)}\in\mathcal{E}^{(t)}$ and $a_{ij}^{(t)}=0$ otherwise. For now, we assume that $A^{(t)}$ is given and provide the average consensus update rule and its convergence property. The construction of $A^{(t)}$ will be illustrated afterwords.

With $A^{(t)}$, to carry out the model aggregation Eq. \eqref{aggregation} in a decentralized manner, each learner $i$ iteratively constructs a sequence of weighted local models $\{\bar\theta_i^{(t)}(k)\}$, where $k$ is the iteration index for the consensus algorithm below, such that $\bar\theta_i^{(t)}(0)=w_i\theta_i^{(t)}$, and the update rule is given by
\begin{align}
\label{consensus update rule}
    \bar\theta_i^{(t)}(k+1)=a_{ii}^{(t)}\bar\theta_i^{(t)}(k)+\sum_{j\in\mathcal{N}_i^{(t)}}a_{ij}^{(t)}\bar\theta_j^{(t)}(k).
\end{align}


For any $k\in\mathbb{N}$, let $\bar\theta^{(t)}(k)=\{\bar\theta_i^{(t)}(k)\}_{i\in\mathcal{V}}$ be the learners' joint state at iteration $k$. Given an initial joint state $\bar\theta^{(t)}(0)$, we say that the learners asymptotically reach average consensus if all the learners' states converge to the average of their initial states as $k$ tends to infinity, i.e.,
\begin{align}
\label{average consensus definition}
    \lim_{k\to\infty}\bar\theta_i^{(t)}(k)=\frac{1}{N}\sum_{j\in\mathcal{V}}\bar\theta_j^{(t)}(0),\;\forall i\in\mathcal{V}.
\end{align}
If Eq. \eqref{average consensus definition} is true, then each learner $i$'s state $\bar\theta_i^{(t)}(k)$ will asymptotically converge to $\bar\theta_i^{(t)}(\infty)=\frac{1}{N}\sum_{j\in\mathcal{V}}\bar\theta_j^{(t)}(0)=\frac{1}{N}\sum_{j\in\mathcal{V}}w_j\theta_j^{(t)}=\frac{1}{N}\theta^{(t)}$, and hence each learner $i$ can derive the global model $\theta^{(t)}$ by computing $N\bar\theta_i^{(t)}(\infty)$.

The following lemma provides a sufficient and necessary condition for reaching average consensus.

\begin{lemma}[\cite{1272421}]
\label{lemma:consensus}
With $A^{(t)}$ in each round $t$ of model aggregation, the learners can achieve asymptotic average consensus Eq. \eqref{average consensus definition} by the update rule Eq. \eqref{consensus update rule} from any initial joint state $\bar\theta^{(t)}(0)$ if and only if the following conditions are simultaneously satisfied
\begin{align}
    &\rho(A^{(t)}-\frac{1}{N}1_N1_N^T)<1,\label{asymptotic convergence}\\
    &1_N^TA^{(t)}=1_N^T,\label{convergence point 1}\\
    &A^{(t)}1_N=1_N,\label{convergence point 2}
\end{align}
where $1_N$ is the $N$-dimensional column vector with all ones, and $\rho(\cdot)$ denotes the spectral radius\footnote{The spectral radius of a square matrix is the largest absolute value of its eigenvalues.} of a square matrix.
\end{lemma}

The intuition of Lemma \ref{lemma:consensus} lies in that condition \eqref{asymptotic convergence} guarantees asymptotic consensus, while conditions \eqref{convergence point 1} and \eqref{convergence point 2} ensure that the convergence is to the desired average point $\frac{1}{N}\sum_{j\in\mathcal{V}}\bar\theta_j^{(t)}(0)$.

The next question is how to construct $A^{(t)}$ that satisfies all the conditions \eqref{asymptotic convergence}--\eqref{convergence point 2}. One efficient approach is the Metropolis-Hastings method, illustrated next.

\noindent\textbf{\emph{Metropolis-Hastings method.}} 
For a time-varying communication graph, the Metropolis-Hastings method \cite{XIAO200733} can be applied to update $A^{(t)}$ to ensure asymptotic average consensus. In particular, for each round $t$, based on its current local communication topology, each learner $i$ constructs weights $a_{ij}^{(t)}$'s for all $j\in\bar{\mathcal{N}}_i^{(t)}$ as follows
\begin{align}
\label{Metropolis-Hastings}
    a_{ij}^{(t)}=\begin{cases}
      \frac{1}{\max\{|\mathcal{N}_i^{(t)}|,|\mathcal{N}_j^{(t)}|\}+1} & \text{if }j\in\mathcal{N}_i^{(t)}\\
      1-\sum\limits_{j\in\mathcal{N}_i^{(t)}}\frac{1}{\max\{|\mathcal{N}_i^{(t)}|,|\mathcal{N}_j^{(t)}|\}+1} & \text{if }j=i,
    \end{cases}
\end{align}
where $|\cdot|$ denotes the cardinality of a set.

\begin{figure}[!ht]
\begin{center}
\includegraphics[width=1\linewidth]{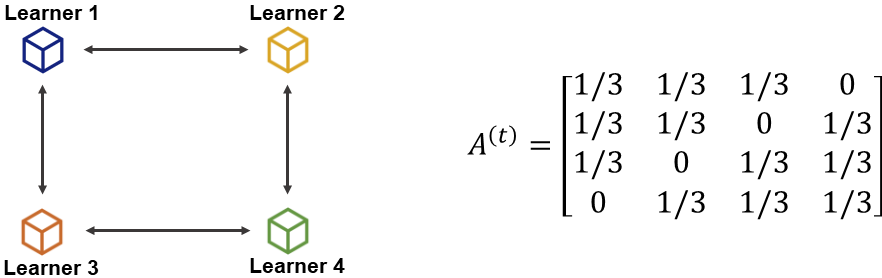}
\caption{An example of matrix $A^{(t)}$ constructed by the Metropolis-Hastings method \eqref{Metropolis-Hastings}.}
\label{doubly_sto_matrix_fig}
\end{center}
\end{figure}

As an illustrative example, in Fig. \ref{doubly_sto_matrix_fig}, the figure on the left shows the communication topology between four learners, and the matrix on the right is the corresponding weighted adjacency matrix $A^{(t)}$ constructed by \eqref{Metropolis-Hastings}.

The following lemma states that $A^{(t)}$ constructed by \eqref{Metropolis-Hastings} satisfies all the conditions of Lemma \ref{lemma:consensus}.

\begin{lemma}[\cite{6854643}]
\label{lemma:Metropolis-Hastings}
Under Assumption \ref{asm:communication graph}, in each round $t$, if $A^{(t)}$ is constructed by \eqref{Metropolis-Hastings}, then the conditions \eqref{asymptotic convergence}--\eqref{convergence point 2} are all satisfied.
\end{lemma}



By Lemmas \ref{lemma:consensus} and \ref{lemma:Metropolis-Hastings}, the update rule \eqref{consensus update rule} with $A^{(t)}$ constructed by \eqref{Metropolis-Hastings} ensures asymptotic average consensus \eqref{average consensus definition}. 
Notice that in running both \eqref{consensus update rule} and \eqref{Metropolis-Hastings}, each learner only needs information from its neighbors. Hence, the aforementioned consensus-based method realizes the model aggregation \eqref{aggregation} over a time-varying communication graph in a fully decentralized manner.

On the other hand, in implementing \eqref{consensus update rule}, each learner $i$ directly sends its state $\bar\theta_i^{(t)}(k)$ at each consensus iteration $k$ to its neighbors. This causes the breach of its initial state $\bar\theta_i^{(t)}(0)$ and of its local training model $\theta_i^{(t)}$. Hence, the privacy issue remains to be addressed for the implementation of \eqref{consensus update rule}.


\subsection{Shamir's Secret Sharing}\label{sec:S4}

In the paper, we will adopt SSS to facilitate privacy in the implementation of \eqref{consensus update rule}. This subsection provides some preliminaries on how to use SSS to distribute a secret over a finite set of entities. More detailed discussions can be found in \cite{Shamir,RC-ID-JBN:2015}.

\noindent\textbf{\emph{Shares generation.}} To distribute a secret $s$ over a set of entities $\mathcal{V}$, SSS uses a polynomial of degree smaller than $|\mathcal{V}|$ to generate $|\mathcal{V}|$ shares of $s$, one share for one entity of $\mathcal{V}$. Formally, given a prime number $p>|\mathcal{V}|$ and a positive integer $\tau<|\mathcal{V}|$, a secret $s\in\mathbb{Z}_p$ is split into $|\mathcal{V}|$ shares $\{\mathcal{H}^i\}_{i\in\mathcal{V}}$ by Algorithm \ref{algo:share generation}. In the algorithm, $p$ is the parameter to set the underlying finite field for SSS, and $\tau$ is the degree of the polynomial used to generate shares of $s$. After shares generation, the share $\mathcal{H}^i$ is sent to entity $i$ for all $i\in\mathcal{V}$.

\begin{algorithm2e}[htbp]
\caption{Shamir's secret shares generation}\label{algo:share generation}

\nonl Syntax: $\{\mathcal{H}^i\}_{i\in\mathcal{V}}={\rm Alg_{ssg}}(s,p,\tau,\mathcal{V})$.

\nonl The executor selects $\tau$ scalars $c_1,\cdots,c_{\tau}\in\mathbb{Z}_p$ uniformly at random with $c_\tau\neq0$, defines a polynomial $\mathcal{H}$ as $\mathcal{H}(\eta)=s+c_1\eta+\cdots+c_{\tau}\eta^{\tau}$, and computes $\mathcal{H}^i=\mathcal{H}(i)\mod p$ for all $i\in\mathcal{V}$.
\end{algorithm2e}

\noindent\textbf{\emph{Reconstruction.}} As given by the following lemma, the secret $s$ can be reconstructed by collecting arbitrary $\tau+1$ or more shares via the technique of Lagrange interpolation. This property directly follows the fact that a polynomial of degree $\tau$ can be uniquely determined by any $\tau+1$ or more points of the polynomial.



\begin{lemma}[\cite{RC-ID-JBN:2015}]
\label{secret reconstruction}
Let $(s,p,\tau,\mathcal{V})$ and $\{\mathcal{H}^i\}_{i\in\mathcal{V}}$ be a set of inputs and corresponding outputs of Algorithm \ref{algo:share generation}, respectively. Then for any set $\mathcal{C}\subseteq\mathcal{V}$ with $|\mathcal{C}|\geq\tau+1$, $s$ can be reconstructed as $s=\sum_{i\in \mathcal{C}}\mathcal{H}^i\delta_{\mathcal{C},i}\mod p$, where
\begin{align}
\label{Lagrange basis polynomial}
    \delta_{\mathcal{C},i}=\prod_{j\in \mathcal{C},j\neq i}\frac{j}{j-i}\mod p,\;\forall i\in\mathcal{C}.
\end{align}
\end{lemma}

\noindent\textbf{\emph{Privacy.}} The privacy property of SSS is given by the following lemma, which states that the collection of any $\tau$ or less shares generated by Algorithm \ref{algo:share generation} contains no information of $s$. This property follows the fact that it takes at least $\tau+1$ points to define a polynomial of degree $\tau$.

\begin{lemma}[\cite{RC-ID-JBN:2015}]
\label{S4 privacy}
SSS provides perfect secrecy against any set $\mathcal{I}\subseteq\mathcal{V}$ such that $|\mathcal{I}|\leq\tau$.
\end{lemma}

\section{New Challenges in Algorithm Design and Analysis}\label{sec:new challenges}

In this section, we first provide the high-level idea of algorithm design based on integrating SSS with average consensus. After that, we identify new challenges to a trivial integration brought by the nature of the concerned problem setting.

\subsection{High-level description}\label{sec:high level description}

As mentioned, in this paper, we achieve privacy-preserving decentralized federated learning by integrating SSS with the average consensus update rule \eqref{consensus update rule}. 
Informally speaking, to protect the privacy of $\bar\theta_i^{(t)}(0)$, each learner $i$ uses a new state $s_i^{(t)}(0)$ as the initial state in executing \eqref{consensus update rule}. The new states $s_i^{(t)}(0)$'s need to simultaneously satisfy:
\begin{itemize}
\item Correctness: The average consensus point under the initial states $s_i^{(t)}(0)$'s is or can be used to locally derive the desired global model $\theta^{(t)}$.
\item Privacy: The observation and the derivation process of $s_i^{(t)}(0)$'s do not disclose any information of $\bar\theta_i^{(t)}(0)$'s.
\end{itemize}

To this end, the learners generate $s_i^{(t)}(0)$'s via SSS, as informally illustrated as follows. First, by Algorithm \ref{algo:share generation}, each learner $i$ uses a polynomial of degree $|\mathcal{N}_i^{(t)}|$ to generate $|\mathcal{N}_i^{(t)}|+1$ shares of $\bar\theta_i^{(t)}(0)$, distributes $|\mathcal{N}_i^{(t)}|$ shares to its corresponding neighbors, while keeping one share private to itself. After the exchange of shares, each learner $i$ aggregates the $|\mathcal{N}_i^{(t)}|$ shares received from its neighbors and the one share generated and held secretly by itself to form $s_i^{(t)}(0)$, and uses it as the initial state in executing \eqref{consensus update rule}.

We next informally discuss the correctness and privacy intuitions of the above procedure.
\begin{itemize}
    \item Correctness: By the convergence property of \eqref{consensus update rule}, all the learners can derive $\sum_{i\in\mathcal{V}}s_i^{(t)}(0)$, which is the aggregation of all the shares of all the learners' local models $\{\bar\theta_i^{(t)}(0)\}_{i\in\mathcal{V}}$. Notice that, by the reconstruction property of SSS, each individual $\bar\theta_i^{(t)}(0)$ can be reconstructed by aggregating all of its $|\mathcal{N}_i^{(t)}|+1$ shares. Hence, $\sum_{i\in\mathcal{V}}s_i^{(t)}(0)$ can be used to reconstruct $\sum_{i\in\mathcal{V}}\bar\theta_i^{(t)}(0)$, which is the global model $\theta^{(t)}$.
    \item Privacy: By the privacy property of SSS, $\bar\theta_i^{(t)}(0)$ is perfectly secret if and only if not all of its $|\mathcal{N}_i^{(t)}|+1$ shares are known to the adversarial learners. It can be perceived that a necessary condition for this is that learner $i$ has at least one benign neighbor.
\end{itemize}


\subsection{New Challenges}\label{subsec:new challenges}

The last subsection presents a high-level framework based on the integration of the consensus method and SSS. However, for the concerned problem setting, a trivial integration is far from enough. In this subsection, we identify new challenges in terms of design and analysis which are critical for establishing rigorous correctness and privacy properties. Besides, we also briefly illustrate how these challenges are addressed in this paper, while the details are provided in Section \ref{sec:algorithm design} and Section \ref{sec:analysis}.

There are four major challenges, as detailed next. Specifically, the first two are due to the real-valued setting of our problem of interest, and bring new challenges to correctness guarantee. The last two stem from more complicated information flow caused by the iterative nature of the consensus process as well as time-varying communication topology between successive training rounds, and bring new challenges to privacy analysis.

\noindent\textbf{\emph{(i) Signed real-valued models.}} The standard SSS scheme involves modular operations and has to be implemented over non-negative integers. However, the training models in federated learning usually take signed real values. To address this mismatch, we propose a transformation between non-negative integers and signed real numbers (given by \eqref{integer to real}). Roughly speaking, the learners transform their local models into integers and apply the procedure described in the last subsection. After the final non-negative integer-valued consensus model is derived, each agent then uses the proposed transformation to turn it back to a signed real-valued model. If the parameter $p$ in Algorithm \ref{algo:share generation} is sufficiently large (a sufficient lower bound of $p$ is provided by \eqref{p bound}), then it is guaranteed that the transformed real-valued model is the correct global model.

\noindent\textbf{\emph{(ii) Termination of consensus iteration.}} As mentioned in the last paragraph, the learners input integer-valued models into the procedure described in Section \ref{sec:high level description}. Hence, theoretically, the asymptotic consensus result is a non-negative integer-valued model. To reconstruct the global model by SSS, before the integer-to-real transformation, this model needs to be exerted a modulo $p$ operation (please refer to Lemma \ref{secret reconstruction} and \eqref{roundness}). However, since the convergence of the consensus update rule is only asymptotic and the weights $a_{ij}^{(t)}$'s in \eqref{consensus update rule} are decimals, the intermediate results of \eqref{consensus update rule} may also be decimals. Due to the subsequent modulo $p$ operation, even if the terminating result is close to the theoretical integer-valued result, there could be a large deviation in the remainder after the modulo operation. To see this, consider the case where the terminating result is 99.4 and rounded to 99, the theoretical result is 100, and the value of $p$ is 50. After the modulo $p$ operation, the remainders for the terminating result and the theoretical result are 49 and 0, respectively. This shows that, compared to usual consensus applications, we need a more careful control on the termination condition. To address this challenge, we identify a sufficient lower bound of the number of consensus iterations (given by \eqref{K bound}) that guarantees that the absolute difference between the terminating and the theoretical results is strictly smaller than 0.5, and hence the result after the rounding operation is just the theoretical result.

\noindent\textbf{\emph{(iii) Privacy leakage during consensus process.}} For the standard SSS-based secure sum computation over a complete communication graph, each entity receives all the other entities' shares just once and then performs an aggregation. For this standard scheme, as long as there is an honest majority (more specifically, the number of adversarial entities is no greater than the degree of the polynomial used to generated shares), then the adversarial entities cannot gain anything beyond the sum of all the entities' private inputs. However, in our case, since the communication graph is sparse, secure sum computation is further facilitated by a consensus process, where the shares need to be iteratively exchanged and aggregated according to the consensus update rule and the underlying communication topology. Such multiple rounds of communications may cause additional privacy leakage, e.g., partial sum (the sum of the local models of a subset of learners). This indicates that new privacy analysis is needed for the consensus process. To address this challenge, we identify a graph-oriented condition, which can be used to characterize the view of the adversarial learners throughout the whole consensus process (please refer to Lemma \ref{lemma:view}).

\noindent\textbf{\emph{(iv) Privacy property under time-varying communication topology.}} Besides the privacy issue caused by the consensus process, the time-varying communication topology further induces new challenges to privacy preservation. Specifically, due to time-varying communication topology, one-shot privacy preservation (privacy for one round of training) is not enough. Instead, we must establish a privacy condition with respect to the evolution of the communication topology. To address this challenge, we further extend the graph-oriented condition mentioned in the last paragraph to derive a sufficient and necessary condition under which perfect secrecy is achieved throughout the evolution of the communication topology (please refer to Theorem \ref{theorem:privacy case i}).

\section{Privacy-Preserving Decentralized Algorithm Design}\label{sec:algorithm design}

In this section, the proposed privacy-preserving decentralized federated learning algorithm is developed. First, we illustrate the design details and highlight how the challenges identified in the last section are addressed. A summarize of the whole design is provided afterwards.

\subsection{Design Details}

In this paper, we use finite precision to cope with transformations between real numbers and integers. In particular, throughout the paper, the precision level is set by $\sigma\in\mathbb{N}$, that is, for any real number, only the first $\sigma$ fraction digits are kept while rest ones are dropped.

The overall design has three phases, secret shares generation of local models, consensus iteration, and global model reconstruction. The design is detailed next.

\noindent\textbf{\emph{Secret shares generation of local models.}} All the learners first agree on a positive prime number $p$, 
which can be realized by a maximum consensus algorithm offline. We next fix a training round $t$ and a learner $i$, and illustrate the secret shares generation of $\bar\theta_i^{(t)}(0)$.

First, for each $j\in\bar{\mathcal{N}}_i^{(t)}$, learner $i$ computes $\delta_{\bar{\mathcal{N}}_i^{(t)},j}^{(t)}$ by \eqref{Lagrange basis polynomial}. Then, for each $l\in\{1,\cdots,n\}$, learner $i$ first transforms $\bar\theta_i^{(t)}(0)$ into an integer via multiplying by $10^\sigma$, and then applies Algorithm \ref{algo:share generation} to use a polynomial of degree $|\mathcal{N}_i^{(t)}|$ to generate $|\bar{\mathcal{N}}_i^{(t)}|=|\mathcal{N}_i^{(t)}|+1$ shares of integer $10^\sigma\bar\theta_{il}^{(t)}(0)$, denoted as $\{\mathcal{H}_{il}^{j(t)}\}_{j\in\bar{\mathcal{N}}_i^{(t)}}={\rm Alg_{ssg}}(10^\sigma\bar\theta_{il}^{(t)}(0),p,|\mathcal{N}_i^{(t)}|,\bar{\mathcal{N}}_i^{(t)})$. In light of Lemma \ref{secret reconstruction}, to facilitate later reconstruction of $10^\sigma\bar\theta_{il}^{(t)}(0)$, learner $i$ further computes $\{\mathcal{S}_{il}^{j(t)}\}_{j\in\bar{\mathcal{N}}_i^{(t)}}$ as
\begin{align}
\label{share generation}
    \mathcal{S}_{il}^{j(t)}=\mathcal{H}_{il}^{j(t)}\delta_{\bar{\mathcal{N}}_i^{(t)},j}^{(t)}\mod p,\;\forall j\in\bar{\mathcal{N}}_i^{(t)}.
\end{align}
For each $j\in\bar{\mathcal{N}}_i^{(t)}$, with $\mathcal{S}_{il}^{j(t)}$ ready for all $l\in\{1,\cdots,n\}$, learner $i$ forms $\mathcal{S}_i^{j(t)}=\{\mathcal{S}_{il}^{j(t)}\}_{l\in\{1,\cdots,n\}}$. Then learner $i$ sends $\mathcal{S}_i^{j(t)}$ to learner $j$ for each $j\in\mathcal{N}_i^{(t)}$, while keeping $\mathcal{S}_i^{i(t)}$ private to itself.


\noindent\textbf{\emph{Consensus iteration.}} The learners agree on a positive integer $K$, which is the number of iterations for running the average consensus algorithm. Again, this can be realized by a maximum consensus algorithm offline. In each training round $t$, upon receiving $\mathcal{S}_{j}^{i(t)}$ generated as above from all of its neighbors $j\in\mathcal{N}_i^{(t)}$, each learner $i$ constructs its new initial state $s_i^{(t)}(0)$ as
\begin{align}
\label{initial state construction}
    s_i^{(t)}(0)=\sum_{j\in\bar{\mathcal{N}}_i^{(t)}}\mathcal{S}_j^{i(t)}\mod p,
\end{align}
and sends $s_i^{(t)}(0)$ to learner $j$ for all $j\in\mathcal{N}_i^{(t)}$. Then, from $k=0$ to $k=K-1$, each learner $i$ iteratively updates its state $s_i^{(t)}(k)$ by
\begin{align}
\label{privacy preserving consensus update rule}
    s_i^{(t)}(k+1)=a_{ii}^{(t)}s_i^{(t)}(k)+\sum_{j\in\mathcal{N}_i^{(t)}}a_{ij}^{(t)}s_i^{(t)}(k),
\end{align}
and sends $s_i^{(t)}(k+1)$ to learner $j$ for all $j\in\mathcal{N}_i^{(t)}$.

\noindent\textbf{\emph{Global model reconstruction.}} At the end of the consensus iteration, each learner $i$ first performs the following roundness\footnote{Given $a\in\mathbb{R}$, denote by $\lfloor a\rfloor$ the greatest integer less than or equal to $a$; by $\lceil a\rceil$ the least integer greater than or equal to $a$; and by $\lfloor a\rceil$ the roundness of $a$, such that $\lfloor a\rceil=\lfloor a\rfloor$ if $a-\lfloor a\rfloor<0.5$, and $\lfloor a\rceil=\lceil a\rceil$ if $\lceil a\rceil-a\leq0.5$.} and modular operations over $s_i^{(t)}(K)$
\begin{align}
\label{roundness}
    z_{il}^{(t)}=\lfloor Ns_{il}^{(t)}(K)\rceil\mod p,\;\forall l\in\{1,\cdots,n\}.
\end{align}
In \eqref{roundness}, the rounding operation is needed to ensure perfect correctness. Specifically, with $A^{(t)}$ generated by \eqref{Metropolis-Hastings}, the update rule \eqref{privacy preserving consensus update rule} ensures that $s_i^{(t)}(k)$ asymptotically converges to the point $\frac{1}{N}\sum_{j\in\mathcal{V}}s_j^{(t)}(0)$. Hence, for each $l\in\{1,\cdots,n\}$, $Ns_{il}^{(t)}(k)$ asymptotically converges to the point $\sum_{j\in\mathcal{V}}s_{jl}^{(t)}(0)$, which is a non-negative integer. However, since the convergence is asymptotic, there could be a difference between $Ns_{il}^{(t)}(K)$ and $\sum_{j\in\mathcal{V}}s_{jl}^{(t)}(0)$. If $K$ is large enough such that the condition $|Ns_{il}^{(t)}(K)-\sum_{j\in\mathcal{V}}s_{jl}^{(t)}(0)|<0.5$ holds, then it is guaranteed that the rounded integer in \eqref{roundness} is equal to the correct consensus point, i.e., $\lfloor Ns_{il}^{(t)}(K)\rceil=\sum_{j\in\mathcal{V}}s_{jl}^{(t)}(0)$. Based on this condition, a sufficient lower bound of $K$ is given by \eqref{K bound} in Section \ref{sec:correctness analysis}.

Notice that each $z_{il}^{(t)}$ is a non-negative integer smaller than $p$ (because it is a remainder of modulo $p$ operation). Each learner $i$ then transforms $z_{il}^{(t)}$ for every $l\in\{1,\cdots,n\}$ back to a signed real number as follows
\begin{align}
\label{integer to real}
\tilde\theta_{il}^{(t)} = \left\{ {\begin{array}{*{20}{l}}
{{z_{il}^{(t)}}/{10^\sigma},\,{\rm if}\,0 \le z_{il}^{(t)} \le {{(p - 1)}}/{2}},\\
{{(z_{il}^{(t)}-p)}/{10^\sigma},\,{\rm if}\, {{(p + 1)}}/{2} \le z_{il}^{(t)} < p}.
\end{array}} \right.
\end{align}
In \eqref{integer to real}, the divide by $10^\sigma$ operation transforms the integer $z_{il}^{(t)}$ into a real number $\tilde\theta_{il}^{(t)}$ with $\sigma$ fraction digits, while the sign of $\tilde\theta_{il}^{(t)}$ is determined by the location of $z_{il}^{(t)}$ in the range of $[0,p)$. For sufficiently large $p$, the sign correctness is guaranteed. Roughly, $p$ needs to be larger than twice of $10^\sigma|\theta_l^{(t)}|$, as informally explained next. Following the reconstruction property of SSS and the convergence of the consensus update rule, we should have $10^\sigma\theta_l^{(t)}\mod p=z_{il}^{(t)}$. The question is, given the remainder $z_{il}^{(t)}$, how to use it to reconstruct $10^\sigma\theta_l^{(t)}$ with the correct sign. Under the condition $p>2\times 10^\sigma|\theta_l^{(t)}|$, if $\theta_l^{(t)}\geq0$, then the remainder $z_{il}^{(t)}$ must locate in the left half of $[0,p)$, while if $\theta_l^{(t)}<0$, then $z_{il}^{(t)}$ must locate in the right half of $[0,p)$. Hence, conversely, as given by \eqref{integer to real}, the location of $z_{il}^{(t)}$ in $[0,p)$ can be used to correctly reconstruct the sign of $\theta_l^{(t)}$. A rigorous sufficient lower bound of $p$ is given by \eqref{p bound} in Section \ref{sec:correctness analysis}.

\begin{algorithm2e}[htbp]
\caption{Privacy-preserving decentralized federated learning}\label{algo:PPDFL}



The learners agree on a positive prime number $p$ and two positive integers $T$ and $K$;

\nonl \ForEach{$i\in\mathcal{V}$}
{Learner $i$ arbitrarily sets $\theta_i^{1,0}\in\mathbb{R}^n$;}

\nonl \For{$t=1$; $t\leq T$; $t=t+1$}
{\nonl \ForEach{$i\in\mathcal{V}$}
{Learner $i$ trains $\theta_i^{(t)}$ by \eqref{local model training};

\nonl \ForEach{$j\in\bar{\mathcal{N}}_i^{(t)}$}
{Learner $i$ constructs $a_{ij}^{(t)}$ by \eqref{Metropolis-Hastings};

Learner $i$ constructs $\delta_{\bar{\mathcal{N}}_i^{(t)},j}^{(t)}$ by \eqref{Lagrange basis polynomial};}
\nonl \ForEach{$l\in\{1,\cdots,n\}$}
{Learner $i$ generates $\{\mathcal{H}_{il}^{j(t)}\}_{j\in\bar{\mathcal{N}}_i^{(t)}}={\rm Alg_{ssg}}(10^\sigma\bar\theta_{il}^{(t)}(0),p,|\mathcal{N}_i^{(t)}|,\bar{\mathcal{N}}_i^{(t)})$ by Algorithm \ref{algo:share generation};

\nonl \ForEach{$j\in\bar{\mathcal{N}}_i^{(t)}$}
{Learner $i$ computes $\mathcal{S}_{il}^{j(t)}$ by \eqref{share generation};}
}
\nonl \ForEach{$j\in\bar{\mathcal{N}}_i^{(t)}$}
{Learner $i$ forms $\mathcal{S}_i^{j(t)}=\{\mathcal{S}_{il}^{j(t)}\}_{l\in\{1,\cdots,n\}}$ and sends $\mathcal{S}_i^{j(t)}$ to learner $j$;}
}
\nonl \ForEach{$i\in\mathcal{V}$}
{Learner $i$ constructs $s_i^{(t)}(0)$ by \eqref{initial state construction} and sends it to learner $j$, $\forall j\in\mathcal{N}_i^{(t)}$;}
\nonl \For{$k=0$; $k<K$; $k=k+1$}
{\nonl \ForEach{$i\in\mathcal{V}$}
{Learner $i$ constructs $s_i^{(t)}(k+1)$ by \eqref{privacy preserving consensus update rule} and sends it to learner $j$, $\forall j\in\mathcal{N}_i^{(t)}$;}
}
\nonl \ForEach{$i\in\mathcal{V}$}
{\nonl \ForEach{$l\in\{1,\cdots,n\}$}
{Learner $i$ constructs $z_{il}^{(t)}$ by \eqref{roundness};

Learner $i$ constructs $\tilde\theta_{il}^{(t)}$ by \eqref{integer to real};}
Learner $i$ forms $\tilde\theta_{i}^{(t)}=\{\tilde\theta_{il}^{(t)}\}_{l\in\{1,\cdots,n\}}$;

Learner $i$ sets $\theta_i^{(t+1,0)}=\tilde\theta_{i}^{(t)}$.
}
}
\end{algorithm2e}

\subsection{Overall Algorithm Design Summary}

Algorithm \ref{algo:PPDFL} presents our overall design, with its operational steps summarized next.

At step 1, all the learners agree on three parameters. In particular, $p$ is the parameter to set the finite field for SSS, $T$ is the number of training rounds, and $K$ is the number of consensus iterations in each training round. As mentioned in the last subsection, these parameters can be realized by a maximum consensus algorithm offline. At step 2, each learner $i$ sets the initial model $\theta_i^{1,0}$ for its local training in the first round. At step 3, each learner $i$ trains its local model $\theta_i^{(t)}$ by $\mathcal{F}_i$ with its initial model $\theta_i^{t,0}$ and dataset $D_i$. At step 4, based on its current local communication topology, each learner $i$ constructs its local weights $a_{ij}^{(t)}$ in $A^{(t)}$ by the Metropolis-Hastings method. At steps 5--8, each learner $i$ applies SSS to generate shares of $10^\sigma\bar\theta_i^{(t)}(0)$ and distributes the shares $\{\mathcal{S}_i^{j(t)}\}_{j\in\mathcal{N}_i^{(t)}}$ to its neighbors. At step 9, each learner $i$ constructs the initial state $s_i^{(t)}(0)$ for the consensus iteration as the sum of all the shares assigned to it. At step 10, each learner $i$ updates its state $s_i^{(t)}(k)$ by the average consensus algorithm with $A^{(t)}$. At steps 11--13, each learner $i$ transforms the consensus model back to a signed real-valued model $\tilde\theta_i^{(t)}$. At step 14, each learner $i$ sets $\tilde\theta_i^{(t)}$ as the initial model $\theta_i^{(t+1,0)}$ for its local training in round $t+1$.

\section{Correctness and Privacy Analysis}\label{sec:analysis}

This section establishes the correctness and privacy properties for Algorithm \ref{algo:PPDFL}.

\subsection{Correctness Analysis}\label{sec:correctness analysis}

The correctness property of Algorithm \ref{algo:PPDFL} is established by the following theorem, which states that each learner $i\in\mathcal{V}$ derives the correct aggregated global model $\theta^{(t)}$ for each round $t$.

\begin{theorem}
\label{theorem:correctness}
Suppose that Assumption \ref{asm:communication graph} holds. By Algorithm \ref{algo:PPDFL}, with sufficiently large $p$ and $K$ such that
\begin{align}
    &p>\max\{N,1+2\times10^{\sigma}N\max_{t,i,l}|\theta_{il}^{(t)}|\},\label{p bound}\\
    &\max_{t}2p\sqrt{N}\|N(A^{(t)})^K-1_N1_N^T\|<1,\label{K bound}
\end{align}
where $\|\cdot\|$ the $\ell_2$ norm of a matrix, it holds that $\tilde\theta_i^{(t)}=\theta^{(t)}$ for all $i\in\mathcal{V}$ and all $t\in\{1,\cdots,T\}$.
\end{theorem}

\textbf{\emph{Proof:}} By Lemma \ref{secret reconstruction}, we have
\begin{align}
\label{correct proof 1}
    \sum_{j\in\bar{\mathcal{N}}_i}\mathcal{S}_{il}^{j{(t)}}\equiv 10^\sigma\bar\theta_{il}^{(t)}(0)\mod p.
\end{align}
By \eqref{initial state construction}, we have
\begin{align}
\label{correct proof 2}
    \sum_{i\in\mathcal{V}}s_{i}^{(t)}(0)\equiv\sum_{i\in\mathcal{V}}\sum_{j\in\bar{\mathcal{N}}_i^{(t)}}\mathcal{S}_{j}^{i{(t)}}\mod p.
\end{align}
Notice that $\sum_{i\in\mathcal{V}}\sum_{j\in\bar{\mathcal{N}}_i^{(t)}}\mathcal{S}_{j}^{i{(t)}}$ is just the sum of all shares generated by all the $N$ learners. Hence, by a rearrangement of the summation order, we have
\begin{align}
\label{correct proof 3}
    \sum_{i\in\mathcal{V}}\sum_{j\in\bar{\mathcal{N}}_i^{(t)}}\mathcal{S}_{j}^{i{(t)}}=\sum_{i\in\mathcal{V}}\sum_{j\in\bar{\mathcal{N}}_i^{(t)}}\mathcal{S}_{i}^{j(t)}.
\end{align}
By \eqref{correct proof 1}, \eqref{correct proof 2} and \eqref{correct proof 3}, we have
\begin{align}
\label{correct proof 4}
    \sum_{i\in\mathcal{V}}s_{i}^{(t)}(0)\equiv\sum_{i\in\mathcal{V}}10^\sigma\bar\theta_{i}^{(t)}(0)\mod p.
\end{align}
Fix any $l\in\{1,\cdots,n\}$. Let $s^{l(t)}(k)=\{s_{il}^{(t)}(k)\}_{i\in\mathcal{V}}$. By \eqref{privacy preserving consensus update rule}, we obtain
\begin{align}
\label{s evolution}
    s^{l(t)}(k+1)=A^{(t)}s^{l(t)}(k),
\end{align}
which further leads to
\begin{align}
\label{s evolution2}
    s^{l(t)}(k)=(A^{(t)})^ks^{l(t)}(0).
\end{align}
Under Assumption \ref{asm:communication graph}, by Lemmas \ref{lemma:consensus} and \ref{lemma:Metropolis-Hastings}, $s_{il}^{(t)}(k)$ asymptotically converges to $\frac{1}{N}1_N^Ts^{l(t)}(0)$, and hence $Ns_{il}^{(t)}(k)$ asymptotically converges to $1_N^Ts^{l(t)}(0)$ for all $i\in\mathcal{V}$. Notice that, for all $i\in\mathcal{V}$, it holds that $0\leq s_{il}^{(t)}(0)<p$, because $s_{il}^{(t)}(0)$ is a remainder of modulo $p$ operation derived by \eqref{initial state construction}. Since $s^{l(t)}(0)$ is an $N$-dimensional vector, we then have
\begin{align}
\label{s bound}
    \|s^{l(t)}(0)\|<p\sqrt{N}.
\end{align}
With a slight abuse of notation, let $A_i^{(t)K}$ be the $i$-th row of $(A^{(t)})^K$. By \eqref{K bound}, \eqref{s evolution2} and \eqref{s bound}, we have
\begin{subequations}
\label{roundedness prepare}
\begin{align}
   &|Ns_{il}^{(t)}(K)-1_N^Ts^{l(t)}(0)|\nonumber\\
   &=|NA_i^{(t)K}s^{l(t)}(0)-1_N^Ts^{l(t)}(0)|\label{roundedness prepare1}\\
   &\leq\|NA_i^{(t)K}-1_N^T\|\|s^{l(t)}(0)\|\label{roundedness prepare2}\\
   &\leq\|N(A^{(t)})^K-1_N1_N^T\|\|s^{l(t)}(0)\|\label{roundedness prepare3}\\
   &<\|N(A^{(t)})^K-1_N1_N^T\|p\sqrt{N}\label{roundedness prepare4}\\
   &<0.5,\label{roundedness prepare5}
\end{align}
\end{subequations}
where the equality \eqref{roundedness prepare1} is due to \eqref{s evolution2}; the inequality \eqref{roundedness prepare2} is a well-known relationship for norm operators; the inequality \eqref{roundedness prepare3} is because $NA_i^{(t)K}-1_N^T$ is the $i$-th row of $N(A^{(t)})^K-1_N1_N^T$, and the $\ell_2$ norm of any one row of a matrix is no greater than that of the whole matrix; the inequality \eqref{roundedness prepare4} is due to \eqref{s bound}; and the inequality \eqref{roundedness prepare5} is due to \eqref{K bound}. Notice that $1_N^Ts^{l(t)}(0)$ is a non-negative integer. By \eqref{roundedness prepare}, we then have $\lfloor Ns_{il}^{(t)}(K)\rceil=1_N^Ts^{l(t)}(0)=\sum_{j\in\mathcal{V}}s_{j}^{(t)}(0)$. By \eqref{correct proof 4}, we then have
\begin{align}
\label{correct proof 5}
    \lfloor Ns_{il}^{(t)}(K)\rceil&\equiv\sum_{j\in\mathcal{V}}10^\sigma\bar\theta_{jl}^{(t)}(0)=\sum_{j\in\mathcal{V}}10^\sigma w_j\theta_{jl}^{(t)}\mod p.
\end{align}
By \eqref{p bound}, noticing that $w_j\leq1$ for all $j\in\mathcal{V}$, we have
\begin{align}
\label{transform prepare}
    p&>1+2\times10^\sigma N\max_{t,i,l}|\theta_{il}^{(t)}|\nonumber\\
    &\geq1+2\times10^\sigma\sum_{j\in\mathcal{V}}|w_j\theta_{jl}^{(t)}|\nonumber\\
    &\geq1+2\times10^\sigma|\sum_{j\in\mathcal{V}}w_j\theta_{jl}^{(t)}|.
\end{align}
By \eqref{transform prepare}, it is either
\begin{align}
\label{case 1}
    0\leq10^\sigma\sum_{j\in\mathcal{V}}w_j\theta_{jl}^{(t)}<(p-1)/2
\end{align}
or
\begin{align}
\label{case 2}
    -(p-1)/2<10^\sigma\sum_{j\in\mathcal{V}}w_j\theta_{jl}^{(t)}<0.
\end{align}
In the case of \eqref{case 1}, by \eqref{roundness} and \eqref{correct proof 5}, we have
\begin{align}
\label{case 1 mod}
    z_{il}^{(t)}=10^\sigma\sum_{j\in\mathcal{V}}w_j\theta_{jl}^{(t)}\mod p=10^\sigma\sum_{j\in\mathcal{V}}w_j\theta_{jl}^{(t)}
\end{align}
By \eqref{case 1} and \eqref{case 1 mod}, we have $0\leq z_{il}^{(t)}<(p-1)/2$. By \eqref{case 1 mod} and \eqref{integer to real}, we then have
\begin{align}
\label{case 1 result}
    \tilde\theta_{il}^{(t)}=z_{il}^{(t)}/10^\sigma=\sum_{j\in\mathcal{V}}w_j\theta_{jl}^{(t)}.
\end{align}
In the case of \eqref{case 2}, by \eqref{roundness} and \eqref{correct proof 5}, we have
\begin{align}
\label{case 2 mod}
    z_{il}^{(t)}=10^\sigma\sum_{j\in\mathcal{V}}w_j\theta_{jl}^{(t)}\mod p=p+10^\sigma\sum_{j\in\mathcal{V}}w_j\theta_{jl}^{(t)}
\end{align}
By \eqref{case 2} and \eqref{case 2 mod}, we have $(p+1)/2<z_{il}^{(t)}<p$. By \eqref{case 2 mod} and \eqref{integer to real}, we then have
\begin{align}
\label{case 2 result}
    \tilde\theta_{il}^{(t)}=(z_{il}^{(t)}-p)/10^\sigma=\sum_{j\in\mathcal{V}}w_j\theta_{jl}^{(t)}.
\end{align}
By \eqref{case 1 result} and \eqref{case 2 result}, we have that $\tilde\theta_{il}^{(t)}=\sum_{j\in\mathcal{V}}w_j\theta_{jl}^{(t)}$ always holds. The above analysis holds for all $t\in\{1,\cdots,T\}$, all $i\in\mathcal{V}$, and all $l\in\{1,\cdots,n\}$. Therefore, by \eqref{aggregation}, we have that $\tilde\theta_i^{(t)}=\theta^{(t)}$ for all $i\in\mathcal{V}$ and all $t\in\{1,\cdots,T\}$. This completes the proof.\hfill\rule{2mm}{2mm}

\begin{remark}
By the analysis above, we can see that perfect average consensus is reached after a finite $K$ number of iterations. We note that this finite average consensus is only due to the usage of finite precision. By \eqref{p bound} and \eqref{K bound}, we can see that the bound of $K$ increases with the value of the precision level $\sigma$. When $\sigma$ tends to infinity, then $K$ also tends to infinity, which indicates asymptotic average consensus.
\end{remark}

\subsection{Privacy Analysis}

To develop the privacy property of Algorithm \ref{algo:PPDFL}, we first introduce the following notions.

Let $\mathcal{B}$ and $\mathcal{A}$ be the sets of benign and adversarial learners, respectively. Notice that $\mathcal{B}\cup\mathcal{A}=\mathcal{V}$. Given any round $t\in\{1,\cdots,T\}$, we say that a subset $\mathcal{B}_s\subseteq\mathcal{B}$ of benign learners are surrounded by $\mathcal{A}$ in $\mathcal{G}^{(t)}$ if there exists a \emph{connected} subgraph of $\mathcal{G}^{(t)}$ consisting of all the benign learners in $\mathcal{B}_s$ but no benign learners in $\mathcal{B}\backslash\mathcal{B}_s$ and no adversarial learners in $\mathcal{A}$, such that for each $i\in\mathcal{B}_s$, it holds that $\mathcal{N}_i^{(t)}\cap(\mathcal{B}\backslash\mathcal{B}_s)=\emptyset$. That is, for every benign learner in $\mathcal{B}_s$, all of its benign neighbors, if any, are inside $\mathcal{B}_s$. Let $\hat{\mathcal{B}}^{(t)}$ be the set containing all such sets $\mathcal{B}_s$'s in round $t$, i.e., $\hat{\mathcal{B}}^{(t)}=\{\mathcal{B}_s\subseteq\mathcal{B}:{\rm the \;learners\; in\;}\mathcal{B}_s{\rm \;are \;surrounded\; by\; }\mathcal{A}{\rm \;in\; }\mathcal{G}^{(t)}\}$.

First, the following lemma establishes the \emph{view} of the adversarial learners throughout the execution of Algorithm \ref{algo:PPDFL}.

\begin{lemma}
\label{lemma:view}
By Algorithm \ref{algo:PPDFL}, in each round $t\in\{1,\cdots,T\}$, the adversarial learners in $\mathcal{A}$ can obtain the value of $\{\sum_{i\in\mathcal{B}_s}\bar\theta_i^{(t)}(0)\}_{\mathcal{B}_s\in\hat{\mathcal{B}}^{(t)}}$, but nothing beyond it.
\end{lemma}

\textbf{\emph{Proof:}} Fix any $t\in\{1,\cdots,T\}$ for concreteness of illustration. Consider any $\mathcal{B}_s\in\hat{\mathcal{B}}^{(t)}$. Let $\bar{\mathcal{B}}_s$ be the complementary set of $\mathcal{B}_s$ in $\mathcal{V}$, i.e., $\bar{\mathcal{B}}_s=\mathcal{V}\backslash\mathcal{B}_s$. For each $l\in\{1,\cdots,n\}$, let $s_{\mathcal{B}_s}^{l(t)}(k)=\{s_{il}^{(t)}(k)\}_{i\in\mathcal{B}_s}$ and $s_{\bar{\mathcal{B}}_s}^{l(t)}(k)=\{s_{il}^{(t)}(k)\}_{i\in\bar{\mathcal{B}}_s}$. With a slight abuse of notation, let $A_{\mathcal{B}_s}^{(t)k}$ be the rows of $(A^{(t)})^k$ corresponding to $s_{\mathcal{B}_s}^{l(t)}(k)$. Moreover, let $A_{\mathcal{B}_s,\mathcal{B}_s}^{(t)k}$ and $A_{\mathcal{B}_s,\bar{\mathcal{B}}_s}^{(t)k}$ be the columns of $A_{\mathcal{B}_s}^{(t)k}$ corresponding to $s_{\mathcal{B}_s}^{l(t)}(k)$ and $s_{\bar{\mathcal{B}}_s}^{l(t)}(k)$, respectively. By \eqref{s evolution2}, we have
\begin{align}
\label{view proof 1}
    s_{\mathcal{B}_s}^{l(t)}(k)&=A_{\mathcal{B}_s}^{(t)k}s^{l(t)}(0)=A_{\mathcal{B}_s,\mathcal{B}_s}^{(t)k}s_{\mathcal{B}_s}^{l(t)}(0)+A_{\mathcal{B}_s,\bar{\mathcal{B}}_s}^{(t)k}s_{\bar{\mathcal{B}}_s}^{l(t)}(0).
\end{align}
By the definition of $\hat{\mathcal{B}}^{(t)}$, for any $i\in\bar{\mathcal{B}}_s$, $s_{il}^{(t)}(0)$ can only reach $s_{\mathcal{B}_s}^{l(t)}(k)$ either directly from or relayed by some learner in $\mathcal{A}$. Therefore, by knowing $A^{(t)}$, the learners in $\mathcal{A}$ can compute the value of $A_{\mathcal{B}_s,\bar{\mathcal{B}}_s}^{(t)k}s_{\bar{\mathcal{B}}_s}^{l(t)}(0)$. For any $i\in\mathcal{B}_s$ such that $\mathcal{N}_i^{(t)}\cap\mathcal{A}\neq\emptyset$, by \eqref{view proof 1}, the learners in $\mathcal{A}$ can derive the value of $A_{i,\mathcal{B}_s}^{(t)k}s_{\mathcal{B}_s}^{l(t)}(0)$ as
\begin{align}
\label{view proof 2}
    A_{i,\mathcal{B}_s}^{(t)k}s_{\mathcal{B}_s}^{l(t)}(0)=s_{il}^{(t)}(k)-A_{i,\bar{\mathcal{B}}_s}^{(t)k}s_{\bar{\mathcal{B}}_s}^{l(t)}(0).
\end{align}
Notice that $A_{i,\mathcal{B}_s}^{(t)k}s_{\mathcal{B}_s}^{l(t)}(0)$ asymptotically converges to $\sum_{i\in\mathcal{B}_s}s_{il}^{(t)}(0)$. This implies that the learners in $\mathcal{A}$ can derive the value of $\sum_{i\in\mathcal{B}_s}s_{il}^{(t)}(0)$. For each $i\in\mathcal{B}_s$, by \eqref{initial state construction} and the definition of $\hat{\mathcal{B}}^{(t)}$, $s_{il}^{(t)}(0)$ can be written as
\begin{align}
\label{view proof 3}
    s_{il}^{(t)}(0)=\sum_{j\in\bar{\mathcal{N}}_i^{(t)}\cap\mathcal{B}_s}\mathcal{S}_{jl}^{i(t)}+\sum_{j\in\mathcal{N}_i^{(t)}\cap\mathcal{A}}\mathcal{S}_{jl}^{i(t)}\mod p.
\end{align}
Notice that in \eqref{view proof 3}, for each $j\in\mathcal{N}_i^{(t)}\cap\mathcal{A}$, $\mathcal{S}_{jl}^{i(t)}$ is generated by the adversarial learner $j$. Hence, the learners in $\mathcal{A}$ know the value of $\sum_{j\in\mathcal{N}_i^{(t)}\cap\mathcal{A}}\mathcal{S}_{jl}^{i(t)}$. By also knowing the value of $\sum_{i\in\mathcal{B}_s}s_{il}^{(t)}(0)$, by \eqref{view proof 3}, the learners in $\mathcal{A}$ can derive
\begin{align}
\label{view proof 4}
    \sum_{i\in\mathcal{B}_s}\sum_{j\in\bar{\mathcal{N}}_i^{(t)}\cap\mathcal{B}_s}\mathcal{S}_{jl}^{i(t)}\equiv\sum_{i\in\mathcal{B}_s}s_{il}^{(t)}(0)-\sum_{i\in\mathcal{B}_s}\sum_{j\in\mathcal{N}_i^{(t)}\cap\mathcal{A}}\mathcal{S}_{jl}^{i(t)}\mod p.
\end{align}
Notice that in \eqref{view proof 4}, $\sum_{i\in\mathcal{B}_s}\sum_{j\in\bar{\mathcal{N}}_i^{(t)}\cap\mathcal{B}_s}\mathcal{S}_{jl}^{i(t)}$ is the sum of all those shares generated by the learners in $\mathcal{B}_s$ that are assigned to the learners in $\mathcal{B}_s$ themselves. For each $i\in\mathcal{B}_s$ and for each $j\in\mathcal{N}_i^{(t)}\cap\mathcal{A}$, $\mathcal{S}_{il}^{j(t)}$ is the share generated by learner $i$ and assigned to the adversarial learner $j$. Hence, the learners in $\mathcal{A}$ know the value of $\sum_{i\in\mathcal{B}_s}\sum_{j\in\mathcal{N}_i^{(t)}\cap\mathcal{A}}\mathcal{S}_{il}^{j(t)}$, which is the sum of all those shares generated by the learners in $\mathcal{B}_s$ that are assigned to the learners in $\mathcal{A}$. Therefore, given the definition of $\hat{\mathcal{B}}^{(t)}$, the learners in $\mathcal{A}$ can derive the sum of all the shares generated by the learners in $\mathcal{B}_s$ as
\begin{align}
\label{view proof 5}
    \sum_{i\in\mathcal{B}_s}\sum_{j\in\bar{\mathcal{N}}_i^{(t)}}\mathcal{S}_{il}^{j(t)}=\sum_{i\in\mathcal{B}_s}\sum_{j\in\bar{\mathcal{N}}_i^{(t)}\cap\mathcal{B}_s}\mathcal{S}_{jl}^{i(t)}+\sum_{i\in\mathcal{B}_s}\sum_{j\in\mathcal{N}_i^{(t)}\cap\mathcal{A}}\mathcal{S}_{il}^{j(t)}.
\end{align}
By the analysis below \eqref{correct proof 3} in the proof of Theorem \ref{theorem:correctness}, we conclude that the learners in $\mathcal{A}$ can then derive the value of $\sum_{i\in\mathcal{B}_s}\bar\theta_{il}^{(t)}(0)$. The above analysis holds for any $t\in\{1,\cdots,T\}$, any $\mathcal{B}_s\in\hat{\mathcal{B}}^{(t)}$ and any $l\in\{1,\cdots,n\}$. Therefore, in each round $t\in\{1,\cdots,T\},$ the adversarial learners in $\mathcal{A}$ can obtain the value of $\{\sum_{i\in\mathcal{B}_s}\bar\theta_i^{(t)}(0)\}_{\mathcal{B}_s\in\hat{\mathcal{B}}^{(t)}}$.

Next we show that the learners in $\mathcal{A}$ do not gain anything beyond the value of $\{\sum_{i\in\mathcal{B}_s}\bar\theta_i^{(t)}(0)\}_{\mathcal{B}_s\in\hat{\mathcal{B}}^{(t)}}$. Let $\mathcal{D}_s\subseteq\mathcal{B}$ be a subset of benign learners that form a connected subgraph within themselves. It suffices to show that if $\mathcal{D}_s$ is not surrounded by $\mathcal{A}$ in $\mathcal{G}^{(t)}$, then the learners in $\mathcal{A}$ do not obtain any information about $\{\bar\theta_i^{(t)}(0)\}_{i\in\mathcal{D}_s}$. Since $\mathcal{D}_s\notin\hat{\mathcal{B}}^{(t)}$, there exists at least one learner $d\in\mathcal{D}_s$ such that $\mathcal{N}_d^{(t)}\cap(\mathcal{B}\backslash\mathcal{D}_s)\neq\emptyset$. Let $d'\in\mathcal{N}_d^{(t)}\cap(\mathcal{B}\backslash\mathcal{D}_s)$. We only need to consider the worst case where $(\mathcal{D}_s\cup\{d'\})\in\hat{\mathcal{B}}^{(t)}$. Similar to the derivation of \eqref{view proof 4}, the learners in $\mathcal{A}$ can derive
\begin{align}
\label{view proof 7}
    \sum_{i\in\mathcal{D}_s\cup\{d'\}}\sum_{j\in\bar{\mathcal{N}}_i^{(t)}\cap(\mathcal{D}_s\cup\{d'\})}\mathcal{S}_{jl}^{i(t)}\equiv\sum_{i\in\mathcal{D}_s\cup\{d'\}}s_{il}^{(t)}(0)-\sum_{i\in\mathcal{D}_s\cup\{d'\}}\sum_{j\in\mathcal{N}_i^{(t)}\cap\mathcal{A}}\mathcal{S}_{jl}^{i(t)}\mod p.
\end{align}
Write the sum $\sum_{i\in\mathcal{D}_s\cup\{d'\}}\sum_{j\in\bar{\mathcal{N}}_i^{(t)}\cap(\mathcal{D}_s\cup\{d'\})}\mathcal{S}_{jl}^{i(t)}$ as
\begin{align}
\label{view proof 8}
    \sum_{i\in\mathcal{D}_s\cup\{d'\}}\sum_{j\in\bar{\mathcal{N}}_i^{(t)}\cap(\mathcal{D}_s\cup\{d'\})}\mathcal{S}_{jl}^{i(t)}=\sum_{i\in\mathcal{D}_s}\sum_{j\in\bar{\mathcal{N}}_i^{(t)}\cap\mathcal{D}_s}\mathcal{S}_{jl}^{i(t)}+\sum_{i\in\mathcal{D}_s\cap\mathcal{N}_{d'}^{(t)}}\mathcal{S}_{il}^{d'(t)}+\sum_{i\in\mathcal{D}_s\cap\mathcal{N}_{d'}^{(t)}}\mathcal{S}_{d'l}^{i(t)}+\mathcal{S}_{d'l}^{d'(t)}.
\end{align}
In the right-hand side of \eqref{view proof 8}, the sum of the first two terms is the sum of all those shares generated by the learners in $\mathcal{D}_s$ that are assigned to the learners in $\mathcal{D}_s$ themselves and to learner $d'$, while the sum of the last two terms is the sum of the shares generated by $d'$ that are assigned to the learners in $\mathcal{D}_s$ and to $d'$ itself. In order to derive the sum $\sum_{i\in\mathcal{D}_s}\bar\theta_{il}(0)$, the learners in $\mathcal{A}$ need to obtain the sum of the first two terms, i.e., $\sum_{i\in\mathcal{D}_s}\sum_{j\in\bar{\mathcal{N}}_i^{(t)}\cap\mathcal{D}_s}\mathcal{S}_{jl}^{i(t)}+\sum_{i\in\mathcal{D}_s\cap\mathcal{N}_{d'}^{(t)}}\mathcal{S}_{il}^{d'(t)}$. By \eqref{view proof 7} and \eqref{view proof 8}, the learners in $\mathcal{A}$ know the value of the modular sum
\begin{align}
\label{view proof 9}
    &\sum_{i\in\mathcal{D}_s}\sum_{j\in\bar{\mathcal{N}}_i^{(t)}\cap\mathcal{D}_s}\mathcal{S}_{jl}^{i(t)}+\sum_{i\in\mathcal{D}_s\cap\mathcal{N}_{d'}^{(t)}}\mathcal{S}_{il}^{d'(t)}+\sum_{i\in\mathcal{D}_s\cap\mathcal{N}_{d'}^{(t)}}\mathcal{S}_{d'l}^{i(t)}+\mathcal{S}_{d'l}^{d'(t)}\mod p.
\end{align}
However, since learner $d'$ is benign, the learners in $\mathcal{A}$ do not know the value of $\sum_{i\in\mathcal{D}_s\cap\mathcal{N}_{d'}^{(t)}}\mathcal{S}_{d'l}^{i(t)}+\mathcal{S}_{d'l}^{d'(t)}$. Since the learners in $\mathcal{A}$ cannot split the modular sum \eqref{view proof 9}, they cannot learn anything about the value of $\sum_{i\in\mathcal{D}_s}\sum_{j\in\bar{\mathcal{N}}_i^{(t)}\cap\mathcal{D}_s}\mathcal{S}_{jl}^{i(t)}+\sum_{i\in\mathcal{D}_s\cap\mathcal{N}_{d'}^{(t)}}\mathcal{S}_{il}^{d'(t)}$. By Lemma \ref{S4 privacy}, this implies that the learners in $\mathcal{A}$ do not gain any information about $\{\bar\theta_i^{(t)}(0)\}_{i\in\mathcal{D}_s}$. This completes the proof.
\hfill\rule{2mm}{2mm}

Based on Lemma \ref{lemma:view}, the perfect secrecy property of Algorithm \ref{algo:PPDFL} is established by the following theorem. It states that the algorithm provides perfect secrecy if and only if all the benign learners in $\mathcal{B}$ form a connected subgraph within themselves for every round $t$. In other words, there is no proper subset of benign learners that are surrounded by $\mathcal{A}$ in any round $t$.

\begin{theorem}
\label{theorem:privacy case i}
Algorithm \ref{algo:PPDFL} provides perfect secrecy against $\mathcal{A}$ if and only if $\hat{\mathcal{B}}^{(t)}=\{\mathcal{B}\}$ for all $t\in\{1,\cdots,T\}$.
\end{theorem}

\textbf{\emph{Proof:}} By Definition \ref{def: perfect secrecy}, if the algorithm provides perfect secrecy against $\mathcal{A}$, then, in each round $t$, the learners in $\mathcal{A}$ must only gain the value of $\sum_{i\in\mathcal{V}}\bar\theta_i^{(t)}(0)$. Notice that the learners in $\mathcal{A}$ know the sum of their own local models, i.e., $\sum_{i\in\mathcal{A}}\bar\theta_i^{(t)}(0)$. Hence, they definitely can infer the sum of all the benign learners' local models $\sum_{i\in\mathcal{B}}\bar\theta_i^{(t)}(0)$ by computing $\sum_{i\in\mathcal{B}}\bar\theta_i^{(t)}(0)=\sum_{i\in\mathcal{V}}\bar\theta_i^{(t)}(0)-\sum_{i\in\mathcal{A}}\bar\theta_i^{(t)}(0)$. Therefore, the algorithm is perfectly secret if and only if the learners in $\mathcal{A}$ do not gain anything about $\{\bar\theta_i^{(t)}(0)\}_{i\in\mathcal{B}}$ beyond the value of $\sum_{i\in\mathcal{B}}\bar\theta_i^{(t)}(0)$ for all $t\in\{1,\cdots,T\}$.

First, if $\hat{\mathcal{B}}^{(t)}\neq\{\mathcal{B}\}$ for some round $t$, then there exists a proper subset $\mathcal{B}_s\subsetneq\mathcal{B}$ of benign learners such that $\mathcal{B}_s\in\hat{\mathcal{B}}^{(t)}$. By Lemma \ref{lemma:view}, the learners in $\mathcal{A}$ can then obtain the value of $\sum_{i\in\mathcal{B}_s}\bar\theta_i^{(t)}(0)$, which is an additional piece of information beyond $\sum_{i\in\mathcal{B}}\bar\theta_i^{(t)}(0)$. Hence, the algorithm is not perfectly secret.

Next, consider the case where $\hat{\mathcal{B}}^{(t)}=\{\mathcal{B}\}$ for all $t\in\{1,\cdots,T\}$. By Lemma \ref{lemma:view}, in each round $t$, the learners in $\mathcal{A}$ gain nothing beyond the value of $\{\sum_{i\in\mathcal{B}_s}\bar\theta_i^{(t)}(0)\}_{\mathcal{B}_s\in\hat{\mathcal{B}}^{(t)}}=\sum_{i\in\mathcal{B}}\bar\theta_i^{(t)}(0)$. Therefore, the algorithm provides perfect secrecy against $\mathcal{A}$. This completes the proof.
\hfill\rule{2mm}{2mm}

\begin{remark}
The condition of Theorem \ref{theorem:privacy case i}, i.e., $\hat{\mathcal{B}}^{(t)}=\{\mathcal{B}\}$ for all $t\in\{1,\cdots,T\}$, ensures the strong privacy property of perfect secrecy such that the adversarial learners in $\mathcal{A}$ do not even know partial sums of the local models of any proper subset of the benign learners. It would be worth noting that, if we only target on the weaker privacy property such that each individual benign learner's local model is not disclosed to the learners in $\mathcal{A}$, then by Lemma \ref{lemma:view}, the condition becomes that each benign learner has at least one benign neighbor in $\mathcal{G}^{(t)}$ for all $t\in\{1,\cdots,T\}$.
\end{remark}





\section{Performance Evaluation}\label{sec:simulation}

This section tests the performance of Algorithm \ref{algo:PPDFL} by a federated learning framework with a real-world dataset.

\subsection{Simulation Setup}

{\bf Environment.} The simulation environment is as follows. On the hardware side, the simulation is performed on a Lenovo ThinkPad laptop computer with Intel(R) Core(TM) i5-1135G7 CPU at 2.40 GHz. On the software side, the simulation is performed on MATLAB R2021b.

{\bf Dataset.} The dataset we use in the simulation is MNIST \cite{6296535}, which is a large-scale dataset of handwritten digits that is broadly used for training various image processing systems. It has a training set of 60000 samples and a testing set of 10000 samples. Each data sample has 784 features and 1 label. The simulation uses the set of 60000 training data samples and evenly distributes them over 100 learners. For each data sample, the label is removed. Hence, each learner has 600 local training data samples and each data sample consists of 784 features.

{\bf ML model for local training.} In each round $t$, each learner $i$ uses an autoencoder to train its local model $\theta_i^{(t)}$. An autoencoder is an unsupervised learning algorithm for neural networks to learn efficient codings of unlabeled data. It has two parts, an encoder that compresses the input into a latent space representation, and a decoder that maps this representation to a reconstruction of the input. With $h$ hidden layers, in each round $t$, each learner $i$'s encoder consists of an $h\times 784$ weight matrix $W_{i(e)}^{(t)}$ and an $h\times 1$ bias vector $b_{i(e)}^{(t)}$, and its decoder consists of a $784\times h$ weight matrix $W_{i(d)}^{(t)}$ and a $784\times 1$ bias vector $b_{i(d)}^{(t)}$. The local model $\theta_i^{(t)}$ is constructed by stacking all the entries of $W_{i(e)}^{(t)}$, $b_{i(e)}^{(t)}$, $W_{i(d)}^{(t)}$ and $b_{i(d)}^{(t)}$ into a single column vector. Therefore, the dimension of $\theta_i^{(t)}$ is $n=h\times 784+h+784\times h+784=1569h+784$.


\subsection{Simulation Results}\label{sec:simulation results}

In the simulation, we set $\sigma=2$, $p=1020431$, and $T=6$. For the correctness verification, we set $K=10$. We have verified that the conditions given by \eqref{p bound} and \eqref{K bound} are both satisfied.

\begin{figure*}[!ht]
\begin{center}
\includegraphics[width=1\linewidth]{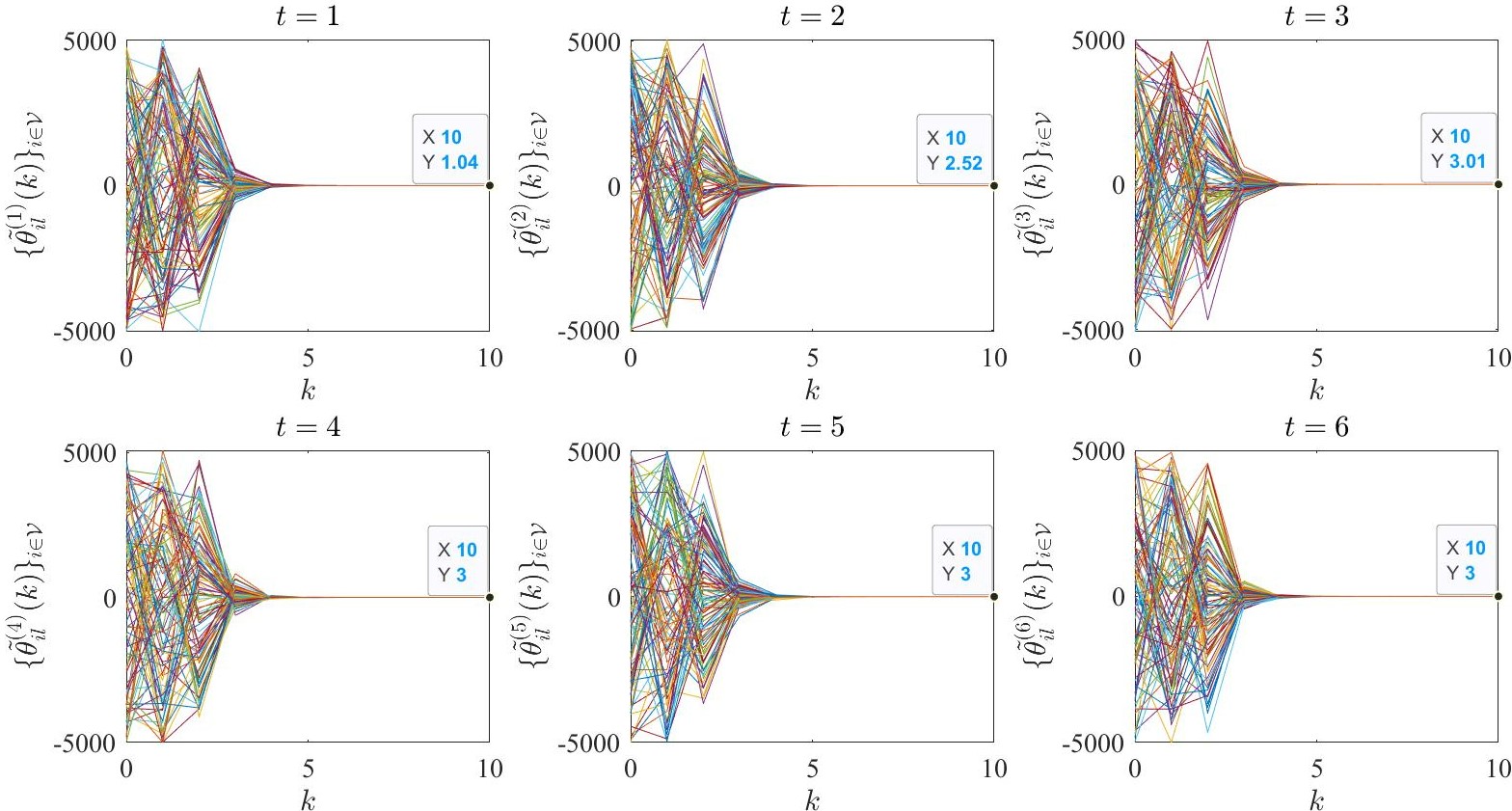}
\caption{Trajectories of $\{\tilde\theta_{il}^{(t)}(k)\}_{i\in\mathcal{V}}$ for $t=1,\cdots,6$, where $\{\tilde\theta_{il}^{(t)}(k)\}$ is the set of $\tilde\theta_{il}^{(t)}(k)$ for $k=0,1,\cdots,10$, and $\{\tilde\theta_{il}^{(t)}(k)\}_{i\in\mathcal{V}}$ is the collection of $\{\tilde\theta_{il}^{(t)}(k)\}$ for all $i\in\mathcal{V}$.}
\label{consensus_fig}
\end{center}
\end{figure*}

\begin{figure}[!ht]
\begin{center}
\includegraphics[width=1\linewidth]{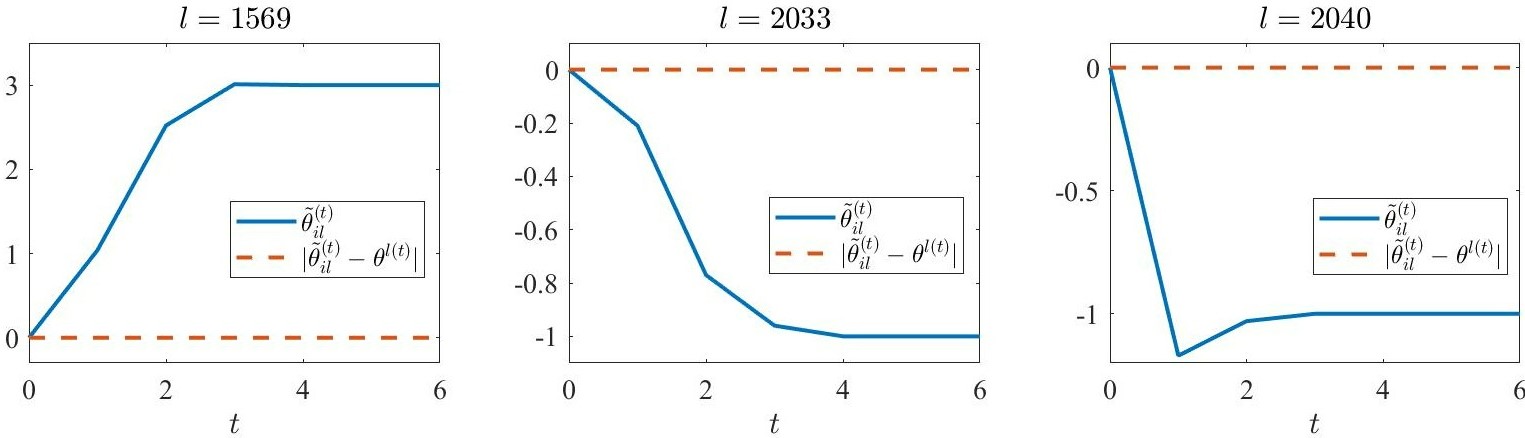}
\caption{Trajectories of $\tilde\theta_{il}^{(t)}$ and $|\tilde\theta_{il}^{(t)}-\theta^{l(t)}|$ for three arbitrarily picked $l$'s.}
\label{global_model_iter}
\end{center}
\end{figure}

We first verify the correctness property of Algorithm \ref{algo:PPDFL}. Here we use one hidden layer for each learner's local training, which leads to $n=2353$. To simulate time-varying communication topology, for each round $t$, an arbitrary communication topology satisfying Assumption \ref{asm:communication graph} is applied. First, we verify correct average consensus at each training round. To this end, we pick an arbitrary $l\in\{1,\cdots,n\}$ for the illustration. In each training round $t$, for each $i\in\mathcal{V}$ and each $k\in\{0,\cdots,K\}$, with $s_{il}^{(t)}(k)$ generated by \eqref{privacy preserving consensus update rule}, we construct $z_{il}^{(t)}(k)$ by \eqref{roundness} with $s_{il}^{(t)}(K)$ replaced by $s_{il}^{(t)}(k)$, and then construct $\tilde\theta_{il}^{(t)}(k)$ by \eqref{integer to real} with $z_{il}^{(t)}$ replaced by $z_{il}^{(t)}(k)$. Notice that $z_{il}^{(t)}=z_{il}^{(t)}(K)$ and $\tilde\theta_{il}^{(t)}=\tilde\theta_{il}^{(t)}(K)$. For each $t=1,\cdots,6$, the trajectories of $\tilde\theta_{il}^{(t)}(k)$ for all $i\in\mathcal{V}$ are sequentially shown in Fig. \ref{consensus_fig}. We can see that, for each round $t$, all the 100 trajectories converge to a same value. Indeed, for each $t=1,\cdots,6$, we have verified that {\bf all the 100 trajectories converge to the correct value of the desired global sum $\theta^{l(t)}=\sum_{i\in\mathcal{V}}w_i\theta_{il}^{(t)}$}, i.e., at $k=K=10$, $\tilde\theta_{il}^{(t)}(10)=\theta^{l(t)}$ for all $i\in\mathcal{V}$. To better show this, we pick an arbitrary $i\in\mathcal{V}$ and plot the trajectories of $\tilde\theta_{il}^{(t)}$ and $|\tilde\theta_{il}^{(t)}-\theta^{l(t)}|$ for three arbitrarily picked $l$'s ($l=1569,2033,2040$); as shown by Fig. \ref{global_model_iter} (notice that these two trajectories are the same for all learners as $\tilde\theta_{il}^{(t)}(k)$'s for all $i\in\mathcal{V}$ converge to a same value). In each sub-figure of Fig. \ref{global_model_iter}, the red dashed curve is the trajectory of $|\tilde\theta_{il}^{(t)}-\theta^{l(t)}|$, i.e., the absolute difference between the consensus value and the ground-truth value at each round $t$. Notice that this curve is constant at 0, which indicates that $\tilde\theta_{il}^{(t)}$ is equal to $\theta^{l(t)}$ at all training rounds. This verifies that the global model computed by Algorithm \ref{algo:PPDFL} is correct under time-varying communication topology. The blue solid curve in Fig. \ref{global_model_iter} is the trajectory of $\tilde\theta_{il}^{(t)}$. It illustrates the convergence of the plain federated learning scheme. 
Fig. \ref{consensus_fig} and Fig. \ref{global_model_iter} together verify the correctness property of Algorithm \ref{algo:PPDFL}.

\begin{figure*}[!ht]
\begin{center}
\includegraphics[width=1\linewidth]{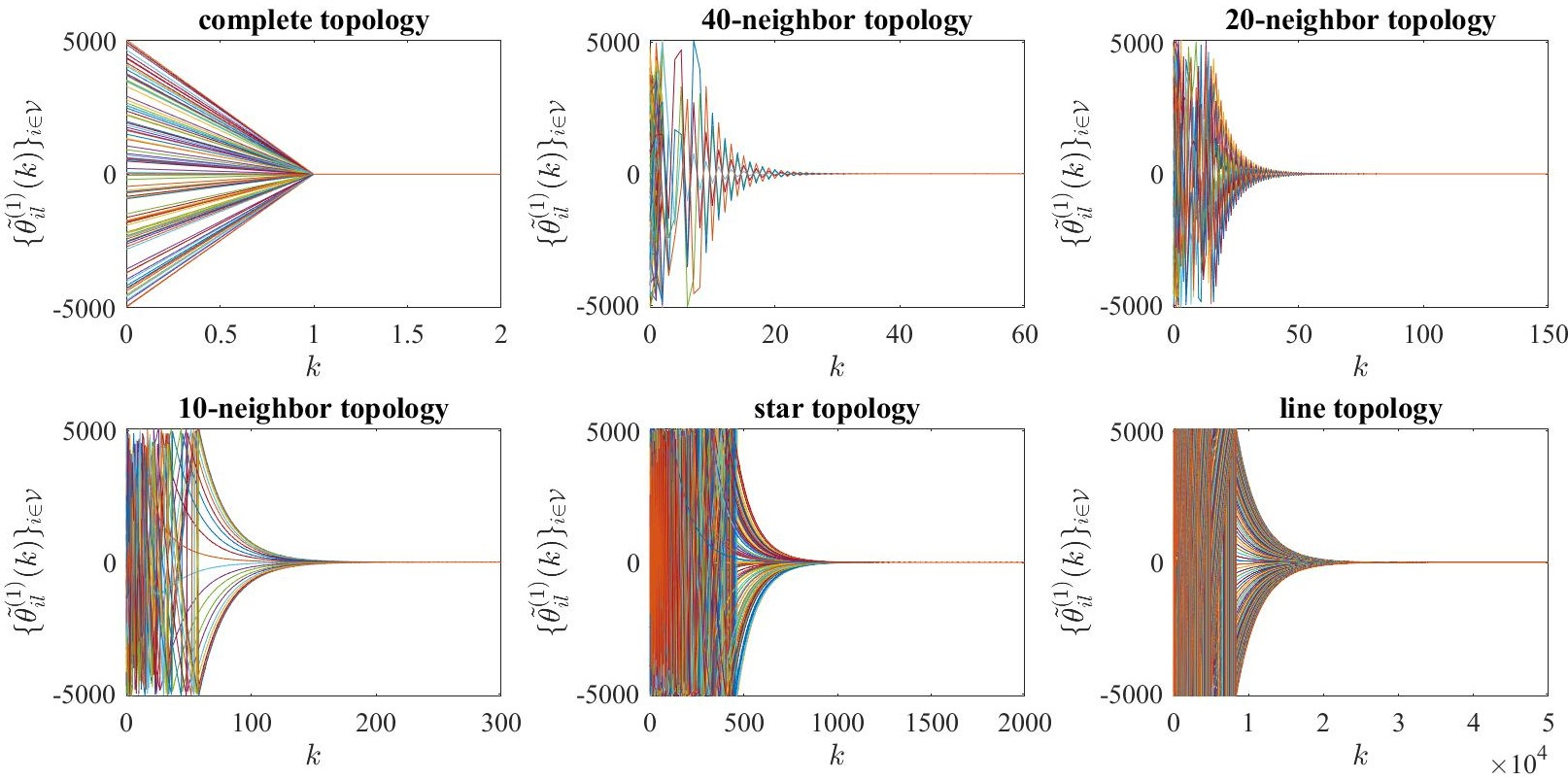}
\caption{Trajectories of $\{\tilde\theta_{il}^{(1)}(k)\}_{i\in\mathcal{V}}$ under different communication topologies.}
\label{consensus_topology_fig}
\end{center}
\end{figure*}

\begin{figure}[!ht]
\begin{center}
\includegraphics[width=1\linewidth]{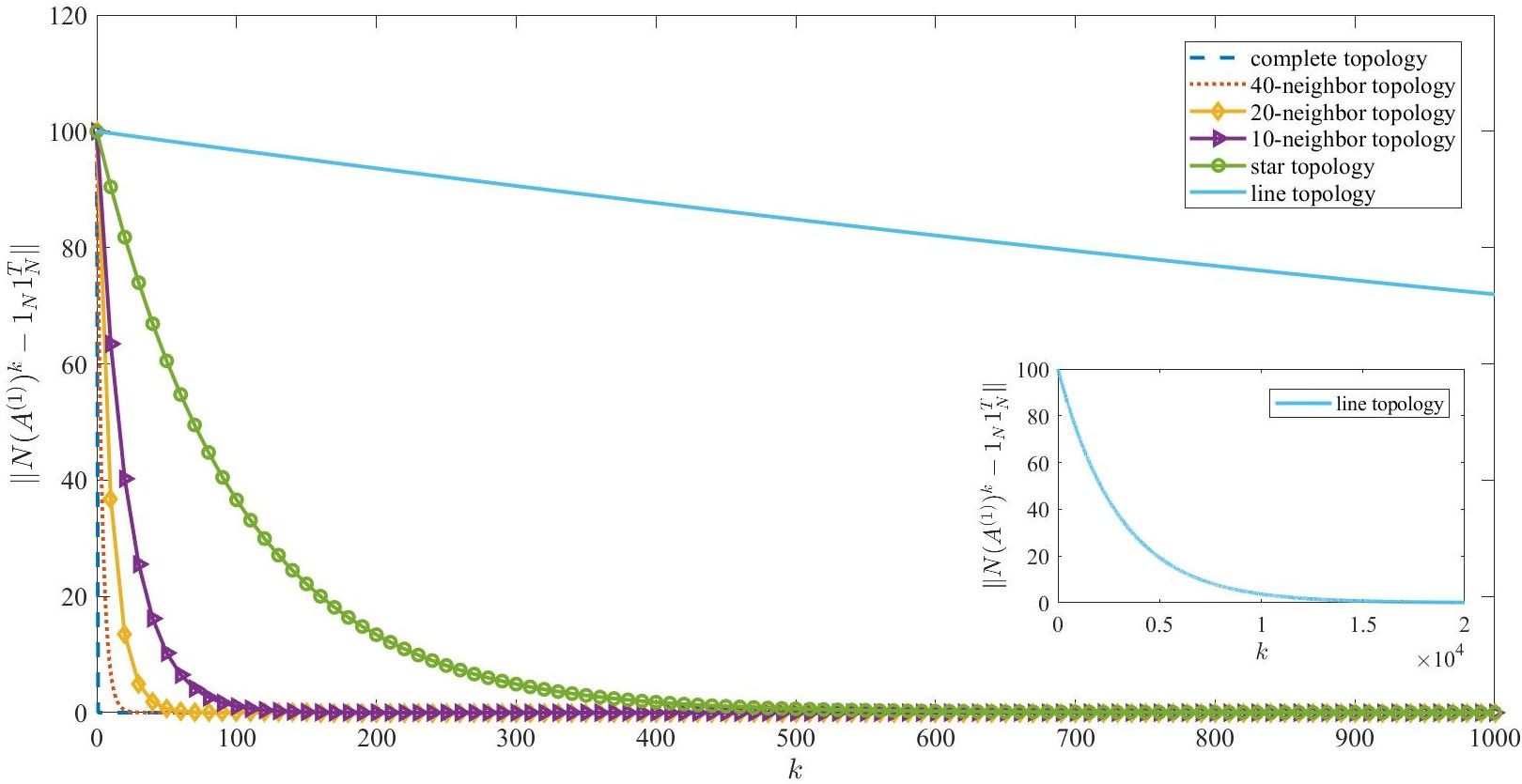}
\caption{Trajectories of $\|N(A^{(1)})^k-1_N1_N^T\|$ under different communication topologies.}
\label{A_matrix_fig}
\end{center}
\end{figure}

We next use simulations to show the impact of the communication topology on the convergence rate of the consensus process. It is well known that the convergence rate is dependent on the overall connectivity degree of the communication topology. Roughly speaking, for a given number of learners, {\bf a denser communication topology usually exhibits a higher convergence rate}. More specifically, the second largest eigenvalue of the underlying weighted adjacency matrix ($A^{(t)}$) is an important indicator of topology connectivity. In our problem setting, a smaller second largest eigenvalue of $A^{(t)}$ indicates a denser connectivity of the communication topology and a better convergence rate \cite{doi:10.1137/060678324}. To visually show the impact of the communication topology on the convergence rate, for the same $l$ as above, we generate the sequence of $\{\tilde\theta_{il}^{(1)}(k)\}_{i\in\mathcal{V}}$ under six communication topologies. The first one is the complete topology, i.e., $(i,j)\in\mathcal{E}^{(1)}$ for all $i,j\in\mathcal{V}$ with $i\neq j$. The second one is a sparse topology where each learner has 40 neighbors. The third and fourth ones are sparser topologies where each learner has 20 and 10 neighbors, respectively. The fifth one is the star topology, i.e., there exists one learner $i$ such that $j\in\mathcal{N}_i^{(1)}$ for all $j\in\mathcal{V}\backslash\{i\}$, while $(j,\ell)\notin\mathcal{E}^{(1)}$ for any $j,\ell\in\mathcal{V}\backslash\{i\}$. The sixth one is the line topology, i.e., the connection of the learners forms a line. Intuitively, the six topologies have descending connectivity degrees. Indeed, their corresponding matrix $A^{(1)}$ have ascending second largest eigenvalues: 0, 0.3259, 0.8181, 0.9555, 0.9900, and 0.9997, respectively. The complete, star and line topologies are three representative communication topologies and have broad applications. In particular, the complete topology has the largest possible connectivity degree (densest) and is typical for, e.g., secure mulitparty computation tasks \cite{RC-ID-JBN:2015}; the star topology depicts the (sparse) spoke–hub distribution paradigm and is common in, e.g., cloud computing \cite{LC:2015}; and the line topology has the smallest possible connectivity degree (sparsest) for connected graphs and is widely used in, e.g., power systems \cite{8340671}. The other three cases depict three different connectivity degrees in between and are used to simulate general sparse graphs covering a wider range of connectivity degrees. The trajectories of $\tilde\theta_{il}^{(1)}(k)$ for all $i\in\mathcal{V}$ under these six communication topologies are shown in Fig. \ref{consensus_topology_fig}. We can see that under all the six communication topologies, all the 100 trajectories converge to the value of the desired global sum 1.04, but clearly with descending convergence rates, which matches discussion above. By \eqref{roundedness prepare}, we can see that the convergence rate can be estimated by the decaying rate of $\|N(A^{(t)})^k-1_N1_N^T\|$. The trajectories of $\|N(A^{(1)})^k-1_N1_N^T\|$ under the above six communication topologies are shown in Fig. \ref{A_matrix_fig} (the small figure shows the convergence under the line topology). It matches the convergence rates observed in Fig. \ref{consensus_topology_fig}. Notice that the trajectory of $\|N(A^k-1_N1_N^T\|$ for a given matrix $A$ can be generated offline. Hence, if the learners have prior knowledge of average connectivity degree of $\mathcal{G}^{(t)}$, then based on the decaying rate of $\|N(A^{(t)})^k-1_N1_N^T\|$ for possible communication topologies, they may be able to choose a less conservative value of $K$.

\begin{figure}[!ht]
\begin{center}
\includegraphics[width=1\linewidth]{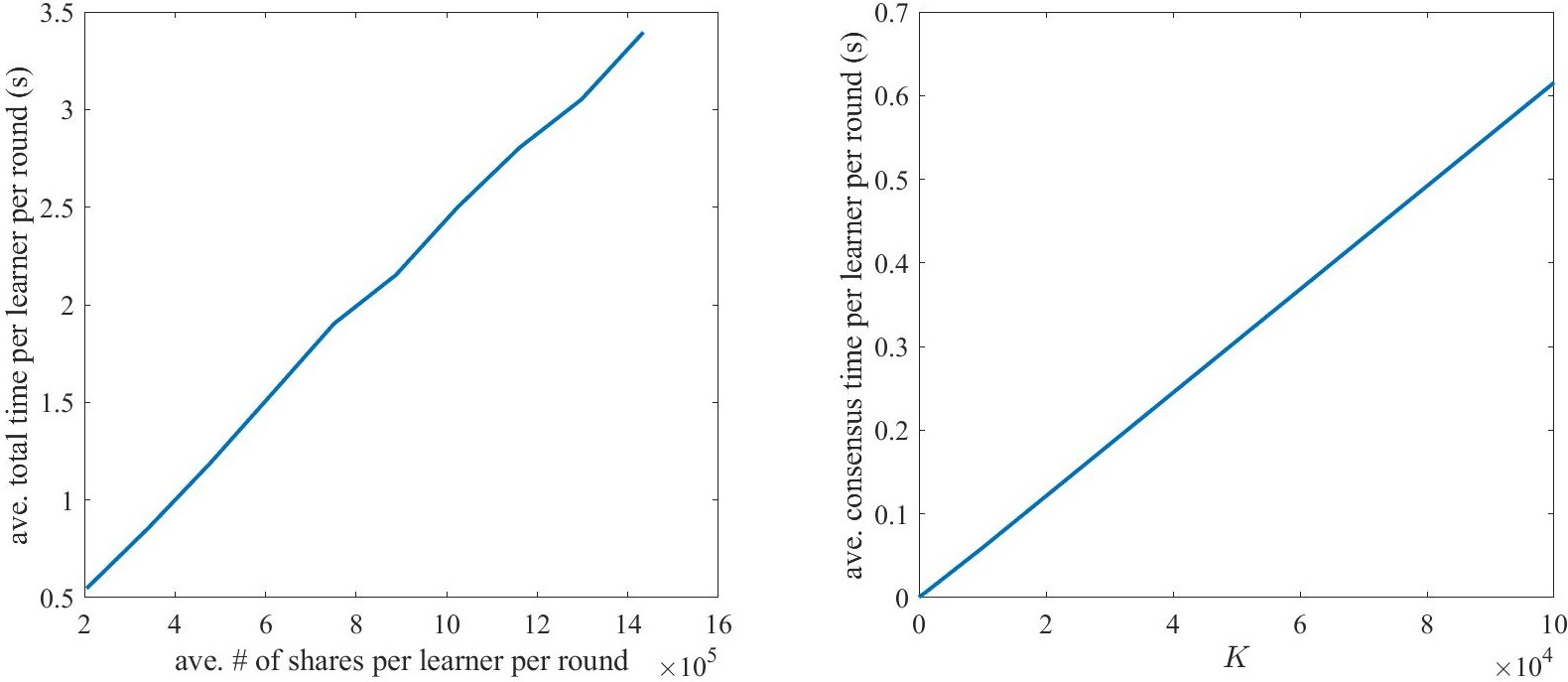}
\caption{Relationship between computational overhead and total number of shares (left) and number of consensus iterations $K$ (right).}
\label{computational_overhead_fig}
\end{center}
\end{figure}

Finally we verify the computational efficiency of Algorithm \ref{algo:PPDFL}. Intuitively, a learner's computational overhead is mainly determined by the total number of shares it needs to generate and the number of iterations of the consensus process. We first examine the relationship between computational overhead and the total number of shares. This number depends on two factors, one is $\bar{\mathcal{N}}_i^{(t)}$, the number of its neighbors including itself, and the other is $n$, the dimension of $\theta^{(t)}$. Notice that the first factor is related to the size of learners. In each training round $t$, a learner's total number of shares is $\bar{\mathcal{N}}_i^{(t)}n$. To examine the relationship between the computational overhead and $\bar{\mathcal{N}}_i^{(t)}n$, an easy way is to tune the values of $n$ by tuning the values of $h$, the number of hidden layers of the learners local autoencoders (recall that $n=1569h+784$). Without loss of generality, the following simulations adopt a fixed communication topology, where the learners' connectivity degrees are well balanced (i.e., they have similar number of neighbors). We use an arbitrary such communication topology where the average number of neighbors for one learner is 87. For each value of $n$, Algorithm \ref{algo:PPDFL} is run for $T=10$ rounds, and in each round $t$, the consensus algorithm is run for $K=10$ iterations. The results are shown in the left sub-figure of Fig. \ref{computational_overhead_fig}, where the $x$-axis is the average number of total shares per learner per round (in this case, $87n$), and the $y$-axis is the average time per learner per training round for the phase of global model aggregation (steps 4--14 of Algorithm \ref{algo:PPDFL}). From the left sub-figure of Fig. \ref{computational_overhead_fig}, we can see that when the average total number of shares per learner per round is $10^6$, the average time per learner per training round is merely around 2.4 seconds. 
In addition, this sub-figure illustrates that the average time per learner per training round grows linearly with the 
average total number of shares per learner per round, where the growth rate is very slow, approximately $2.315\times 10^{-6}$. This verifies that {\bf our algorithm is computationally efficient and scales well with large-size dense networks and high dimensional training models}. Next we examine the relationship between the running time for the consensus process (step 10 of Algorithm \ref{algo:PPDFL}) and the number of consensus iterations $K$. The above communication topology is adopted and fixed. For each value of $K$, Algorithm \ref{algo:PPDFL} is run for $T=10$ rounds, and in each round $t$, the consensus algorithm is run for $K$ iterations. The results are shown in the right sub-figure of Fig. \ref{computational_overhead_fig}, where the $x$-axis is the value of $K$, and the $y$-axis is the average consensus time per learner per training round. From the right sub-figure of Fig. \ref{computational_overhead_fig}, we can see that when $K=10^5$, the average consensus time per learner per training round is merely around 0.6153 seconds. In addition, this sub-figure shows that the average consensus time per learner per training round grows linearly with the 
value of $K$, where the growth rate is also very slow, approximately $6.155\times 10^{-6}$. This verifies that {\bf our algorithm is also computationally efficient for sparse networks which may need a large number of consensus iterations}.


\section{Conclusions and Future Works}\label{sec:conclusion}

This paper developed a new algorithm for privacy-preserving decentralized federated learning over a time-varying communication graph. A consensus-based framework is adopted to enable decentralized global model aggregation. In each round of model aggregation, the Metropolis-Hastings method is applied to update the weighted adjacency matrix based on the current communication topology so as to ensure convergence of average consensus. The technique of Shamir's secret sharing scheme is further integrated with the consensus-based framework to facilitate privacy preservation. The correctness and privacy properties of the proposed algorithm are both analyzed. Its correctness, convergence rate and computational overhead are examined by a case study on a federated learning application using the MNIST dataset. Beyond global model aggregation in federated learning, the proposed algorithm can be readily applied to general secure aggregation tasks over sparse time-varying communication graphs, e.g., decentralized opinions agreement, multi-vehicle rendezvous, and energy supply/consumption aggregation. Moreover, it can also be applied to facilitate privacy for more complicated problems that are solvable by consensus-based approaches \cite{8716798}, e.g., distributed formation control, state estimation, unconstrained convex optimization, and resource allocation.

An interesting future work topic is to extend the attacker model to also include active attacks. For example, with external data poisoning attacks, data transmitted over communication links may be tampered by external attackers; and with Byzantine attacks, the learners themselves may be corrupted to maliciously deviate from the designed algorithm. In the presence of both passive and active attacks, a resilient algorithm needs to be developed which can preserve privacy and meanwhile maintain a satisfactory learning performance.


\bibliographystyle{ACM-Reference-Format}
\bibliography{PPDFL}
\end{document}